\documentclass[a4paper,11pt]{article}
\pdfoutput=1 

\usepackage{jheppub}

\usepackage[utf8]{inputenc}
\usepackage{amsmath}
\usepackage{float}
\usepackage{amssymb}
\usepackage{graphicx}
\usepackage{multirow}
\usepackage[normalem]{ulem}
\usepackage{comment}
\usepackage{slashed}
\usepackage{mathtools}
\usepackage{pifont}
\usepackage[table,xcdraw,dvipsnames]{xcolor}

\usepackage{tikz-feynman}
\tikzfeynmanset{compat=1.1.0}
\usetikzlibrary{arrows,automata,positioning}
\tikzset{
	vertex/.style={circle,draw, minimum size=1.5em},
	edge/.style={->,> = latex'}
}

\definecolor{darkgreen}{rgb}{0.0, 0.5, 0.0}  
\newcommand{\med}[1]{\langle #1\rangle}

\newcommand{\braket}[1]{\ensuremath{\left\langle #1 \right\rangle}}

\newcommand{\mpl}{M_{\rm P}}

\newdimen\arrowsize
\newdimen\mylw
\pgfkeys{/my arrows/chemeq/.style={draw,thick,double distance=2pt,onearc-onearc}}
\pgfkeys{/my arrows/size/.code={\pgfsetarrowoptions{onearc}{#1}}}
\def\myalw{.4pt}
\pgfarrowsdeclare{onearc}{onearc}{
	\mylw=\myalw
	\pgfarrowsleftextend{-\pgfgetarrowoptions{onearc}-.5\mylw}
	\pgfarrowsrightextend{0pt}
}{
	\pgfsetdash{}{0pt}
	\mylw=\pgflinewidth
	\pgfsetlinewidth{\myalw}
	\advance\arrowsize by.5\pgflinewidth
	\pgfpathmoveto{\pgfpoint{-\pgfgetarrowoptions{onearc}}{-\pgfgetarrowoptions{onearc}-.5\mylw}}%
	\pgfpatharc{180}{90}{\pgfgetarrowoptions{onearc}}
	\pgfusepathqstroke
}

\title{\boldmath 
    Dark Matter in an Evanescent Three-Brane Randall-Sundrum Scenario
}

\author[b]{Andrea Donini,\,}
\author[b]{Miguel G. Folgado,\,}
\author[a,b]{Juan Herrero-Garc\'{\i}a,\,}
\author[b]{Giacomo Landini,\,}
\author[a,b]{Alejandro Muñoz-Ovalle}
\author[a,b]{,\,and Nuria Rius}

\affiliation[a]{Departament de Física Teòrica, Universitat de València, 46100 Burjassot, Spain}

\affiliation[b]{Instituto de Física Corpuscular (CSIC-Universitat de València),
Parc Científic UV, C/Catedrático José Beltrán, 2, E-46980 Paterna, Spain}

\emailAdd{donini@ific.uv.es}
\emailAdd{migarfol@ific.uv.es}
\emailAdd{juan.herrero@ific.uv.es}
\emailAdd{giacomo.landini@ific.uv.es}
\emailAdd{almuo@ific.uv.es}
\emailAdd{nuria.rius@ific.uv.es}

\abstract{
Apart from its gravitational interactions, dark matter (DM) has remained so far elusive in laboratory searches.  One possible explanation is that the relevant interactions to explain its relic abundance are mainly gravitational. In this work we consider an extra-dimensional Randall-Sundrum scenario with a TeV-PeV IR brane, where the Standard Model is located, and a GeV-TeV  deep IR (DIR) one, where the DM lies. When the curvatures of the bulk to the left and right of the IR brane are very similar, the tension of the IR brane is significantly smaller than that of the other two branes, and therefore we term it ``evanescent". In this setup, the relic abundance of DM arises from the freeze-out mechanism, thanks to DM annihilations into radions and gravitons. Focusing on a scalar singlet DM candidate, we compute and apply current and future constraints from direct, indirect and collider-based searches. Our findings demonstrate the viability of this scenario and highlight its potential testability in upcoming experiments. We also discuss the possibility of inferring the number of branes if the radion and several Kaluza-Klein graviton resonances
are detected at a future collider.
}


\begin{document}
	\maketitle

\section{Introduction} \label{sec:intro}

	The nature of dark matter (DM) remains one of the open issues of the Standard Model (SM). Its existence has been inferred purely from gravitational interactions. Excluding the latter, the quest for detecting it in laboratory experiments has proven elusive thus far. However, if DM exclusively interacts via the usual gravitational interactions, its relic abundance cannot stem from thermal freeze-out. On the other hand, several beyond the SM scenarios incorporate extra spatial dimensions in order to solve the electroweak hierarchy problem and/or to provide a theory of quantum gravity, such as string theory. In these setups, once the extra-dimensions are compactified, heavy Kaluza-Klein (KK) modes of the gravitons emerge. These have enhanced interactions with matter compared to the conventional gravitational ones mediated by the zero mode. Moreover, a stable brane demands the presence of a new scalar degree of freedom (dof), the radion, which also has significant interactions. This raises an intriguing question: 
	\vspace{0.3cm}
    
	\emph{Could these intensified gravitational interactions in extra-dimensional theories potentially facilitate the generation of the DM relic abundance via thermal freeze-out?}
    
	\vspace{0.3cm}

\begin{figure}[htbp]
	\centering
\includegraphics[width=0.5\textwidth]{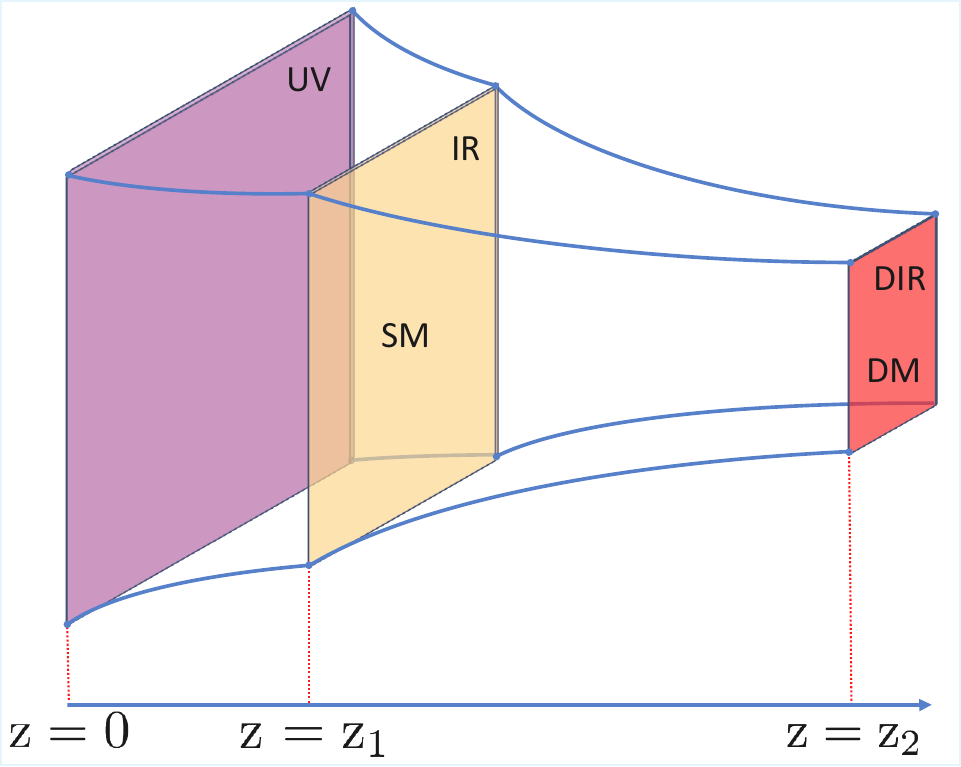}
 	\caption{Schematic representation of the three-brane setup used in this work in conformal coordinates. }\label{fig:3branes}
\end{figure}

To explore it, in the following we consider a Randall-Sundrum (RS) framework as a well-motivated solution capable of addressing the SM hierarchy problem \cite{RandallLisa1999LMHf,RandallLisa1999AAtC}. It involves three branes: the UV Planck-scale brane at conformal coordinate $z=0$, the infrared (IR)
 intermediate brane where the SM lives, located at $z_1\sim {\rm PeV^{-1}-TeV^{-1}}$, and the deep IR (DIR)
 brane where the DM resides, at $z_2\sim {\rm  TeV^{-1}-GeV^{-1}}$. In Fig.~\ref{fig:3branes} we illustrate the three-brane framework considered in this work.
 
In this setup, graviton-DM interactions are enhanced with respect to SM-graviton or graviton-mediated SM-DM interactions. This three-brane setup has been studied in Refs.~\cite{Seung1,Seung2}. When the curvatures of the bulk to the left and to the right of the IR brane are very similar, $k_2 \sim k_1$, the brane tension of the IR brane vanishes, thus making it {\em evanescent}. In this limit, 
computations are greatly simplified. We consider, therefore, the scenario in which an
evanescent brane with the SM fields is located at $z_1$, 
and a DIR brane, where the DM remains, is located at $z_2$. To stabilise these branes via the Goldberger-Wise mechanism necessarily requires the existence of two radion modes. Within the evanescent brane perspective, though, one of the two radions basically decouples, whereas the second, lighter one, drives the DM phenomenology, as in standard two-brane RS scenarios. In Ref.~\cite{Donini:2025qrf}, some of us study the generalization to the case in which $k_1\neq k_2$.
 
We consider a real scalar singlet DM candidate, but the results are not expected to change significantly 
as long as the DM annihilation cross-section is $s-$wave (see, for example, what happens in the two-brane case, Ref.~\cite{Brax:2019koq,Folgado:2020vjb,Chivukula:2024nzt}). 
We assume that the DM particle is stable. This can be realised, for instance, if we postulate a $Z_2$ symmetry under which the DM is odd and the SM and bulk particles are even. In our scenario, the usual Higgs portal is forbidden and therefore DM primarily undergoes annihilations into gravitons/radions\footnote{The latter, as we will show, decay into SM well before the onset of Big Bang Nucleosynthesis (BBN).}. This differentiates our work from the previous publication \cite{Folgado:2019sgz}, where DM dominantly annihilates into SM particles (through a graviton/radion exchange) and the DM relic abundance is reproduced in a different region of the parameter space. The configuration studied in this paper provides direct detection (DD), indirect detection (ID) and collider signals, and their study constitutes the main objective of our investigation. We consider a scenario in which the interactions among DM, bulk particles (KK gravitons and the radion) and SM particles are strong enough to keep the three in thermal equilibrium in the early Universe, so that the DM does not belong to a secluded sector.

Various related works are available in the literature. In Ref.~\cite{Lee} the authors study an effective two-brane scheme in the context of conformal field theory, where the dilaton plays the role of the radion. They consider a GeV-scale DIR brane, motivated by the stochastic gravitational wave background detected by pulsar timing array (PTA) experiments \cite{NANOGrav:2023hvm,Xu_2023,Reardon_2023,2023}, which can be generated by a first order phase transition triggered by the DIR brane. To reconcile the relic abundance of GeV-scale DM and abide by constraints from ID, they resort to forbidden annihilations of DM into radions.
Reference~\cite{Koutroulis:2024wjl} has the same three-brane setup as us, but DM is instead a fermion, which annihilates only into radions since the channels into KK gravitons are kinematically closed. Furthermore, the authors fix the DM mass in the range $m_{\rm DM} \simeq [10^{-2} - 10]$ GeV and the radion mass in the range $m_{r} \simeq [10^{-3} - 10^3]$ MeV, in order to explain also the PTA results. Such mass ranges are allowed by ID bounds, because annihilations are $p$-wave suppressed. This is a characteristic feature of a fermionic DM annihilating to scalar particles (the radion). However, this is not the case if the DM annihilation to KK gravitons ($s$-wave) is kinematically allowed. In our setup, we study scalar DM and consider a wider range of masses and DIR scales, including annihilations into KK gravitons in the evanescent three-brane RS setup. 

In Ref.~\cite{Chivukula_2020} the authors study a two-brane setup with a range of $m_{\rm DM}$ similar to ours. They study different values for $\Lambda_{\rm IR}\simeq [20 - 80]$ TeV and find that the relic abundance cannot be reproduced for scalar DM. They also find that for fermionic DM, $\Lambda_{\rm IR}\simeq [20 - 40]$ TeV is needed, while for vector DM, the relic abundance is obtained for any of the values within the range of study. Although the values of $\Lambda_{\rm IR}$ considered resemble those of our $\Lambda_{\rm DIR}$, the relic abundance is achieved through DM annihilations into SM particles, whereas in our three-brane setup annihilations into bulk particles dominate. Finally, Ref.~\cite{Cacciapaglia} discusses the same three-brane setup where the radion is identified with the DM. Summarising, the main novelty of our work consists in the inclusion of DM annihilations into gravitons in this scenario and performing a comprehensive study of the phenomenology.

Additionally, it is essential to highlight that in the original publications \cite{Folgado:2020vjb,Folgado:2019sgz, Cacciapaglia, Gill:2023kyz} certain diagrams relevant for the computation of DM annihilations were missing, causing some cross-sections with KK gravitons or radions in the propagators to violate unitarity stronger than expected, well below the effective Planck scale. However, IR physics such as compactification should not have any other cut-off but the latter. The solution to this problem has been showcased in subsequent references \cite{Chivukula_2020, 2206.10628,de_Giorgi_2021,deGiorgi:2021xvm, Chivukula:2023sua, deGiorgi:2023mdy,Chivukula:2024nzt}, employing various methodologies such as sum rules \cite{Chivukula_2019,de_Giorgi_2021,deGiorgi:2021xvm,Chivukula:2022kju}, exhaustive computational efforts \cite{Chivukula_2020b}, Ward-identities \cite{Gill_2023}, computations in the 't Hooft gauge \cite{Chivukula:2023qrt}, etc. In particular, using Goldstone's theorem one can see that the unphysical KK graviphotons and KK graviscalars cure the high-energy behaviour, similarly to the SM situation with the longitudinal $W$-boson amplitudes and the Higgs boson. These unphysical modes are \emph{Higgsed} by the KK gravitons, which then become massive. We have verified the validity of these \emph{miraculous} cancellations in the present analysis by performing the brute force computation, including the whole tower of KK gravitons.\footnote{In our numerical computations we include up to the first $\mathcal{O}(100)$ KK gravitons. We have checked that including more states does not modify the results.}
As a result, the theory still violates unitarity (as expected by an effective quantum theory of gravity), but with a softer dependence on $s$.\footnote{Note that computing cross-sections that include couplings of KK gravitons with fields on both the IR brane and in the bulk, such as in the case ${\rm SM\, SM} \rightarrow G_n G_m$, requires a full three-brane setup, with two propagating radions, to canonically normalise the Lagrangian, and so on.}

The subsequent sections of this manuscript are organised as follows. In Sec.~\ref{sec:setup} we describe the RS framework considered in this work. In Sec.~\ref{sec:decays} we discuss the decay rates of the radion and KK gravitons. In Sec.~\ref{sec:relic} we analyse the different contributions to the relic abundance from thermal freeze-out. In Sec.~\ref{sec:bounds} we study the constraints on the model. The main results of the paper are presented in Sec.~\ref{sec:results}, and our conclusions are summarised in Sec. \ref{sec:conc}. 

We also include several appendices with additional content. 
In App.~\ref{app:Int} we provide the relevant interactions, in App.~\ref{app:DD} we discuss DM direct detection, and in App.~\ref{app:decayrates} we provide the decay widths of the radion and the KK gravitons.

\section{An evanescent three-brane Randall-Sundrum scenario} \label{sec:setup}

In this Section we present the theoretical framework whose 
phenomenological implications are the main goal of this paper. In Sec.~\ref{sect:twobranes} we briefly review the standard two-brane RS scenario. After that, in Sec.~\ref{sect:threebranes} we introduce the so-called ``+ + -" three-brane warped model that we are going to study, taking advantage of existing literature. 
	
\subsection{A short review of the standard Randall-Sundrum scenario}
\label{sect:twobranes}	
	
The popular RS scenario \cite{RandallLisa1999LMHf} considers a non-factorizable 5-dimensional metric in the form
\begin{equation}
ds^2 = \bar g^{(5)}_{MN} \,  dx^M dx^N  = e^{- 2 \sigma} \, \eta_{\mu\nu} \, dx^\mu \, dx^\nu - dy^2,
\end{equation}
where $M,N=0,\ldots, 4$ and $\bar g^{(5)}_{MN}$ is the metric, $\sigma = k |y|$, with $k$ the curvature along the 5-th dimension, and the signature of the metric is $(+,-,-,-,-)$. Here $\mu,\nu=0,\ldots, 3$ and $\eta_{\mu\nu}$ is 4-dimensional metric. The extra dimension $y$ is compactified on a circle of radius $r_c$. A $Z_2$ reflection symmetry is also imposed, such that two points (traditionally, $y = 0$ and $y = \pi r_c$) are singular. Two branes are considered (the UV and the IR brane), located at these two singular points. The ``standard" RS setup assumes the UV-brane at $y = 0$ and the IR-brane (where the SM is usually located) at $y = \pi r_c$. Within this particular choice, all mass scales, such as $k$ or the fundamental scale of gravity $M_5$, are of the order of the Planck mass, $M_{\rm P}\simeq1.2\times 10^{19}$ GeV. The action in 5D is
\begin{equation}
    S_\text{RS}= - M_5^3\,\int d^{4}x \int_0^{\pi r_c} dy\, \sqrt{g^{(5)}} \, \left[ R^{\left(5\right)} + 2 \, \Lambda_{5} \right], \label{eq:RSaction}
\end{equation}
where $R^{\left(5\right)}$ is the 5D Ricci scalar, 
$g ^{(5)}$ is the determinant of the 5D metric $g^{(5)}_{MN}$ given above and $\Lambda_5$ is the 5D cosmological constant.
By solving Einstein's equations of General Relativity, the following relation between the 5D cosmological constant $\Lambda_{5}$ and the curvature $k$ may be obtained \cite{RandallLisa1999LMHf}:
\begin{equation} \label{k curvature}
    k=\sqrt{-\frac{\Lambda_{5}}{6}} \, ,
\end{equation}
from which we see that this scenario has the bulk geometry of an anti-de Sitter 5D spacetime, with negative cosmological constant $\Lambda_5$.
within this setup, the relation between the Planck mass in 4D and the fundamental scale in 5D, $M_5$, is
\begin{equation}
    \label{eq:MplanckRS}
    \bar M_{\rm P}^2 = \frac{M_5^3}{k} \, \left (1 -e^{-2 \pi k r_c} \right ),
\end{equation}
where $\bar M_{\rm P}$ is the reduced Planck mass, $\bar M_{\rm P} = M_{\rm P}/\sqrt{8 \pi}$. For a different choice of the location of IR and UV branes, see App.~C of Ref.~\cite{Giudice:2016yja}.

In order to have a stable anti-de Sitter background metric on the segment $y \in [0,\pi r_c]$ it is mandatory  to introduce additional terms in the action 
localized at the fixed points of the orbifold, 
$y = 0$ and $y = \pi r_c$: 
\begin{equation}
    \label{eq:RSbraneterms}
    S_{\rm branes} = \sum_{i = \rm UV, \rm IR} 
    \int d^4 x \int_0^{\pi r_c} dy \sqrt{-g_i^{(4)}} \delta (y - y_i) 
    \left \{ - s_i + \dots \right \} \, , 
\end{equation}
where $s_i$ are the brane tensions and dots refer to the Lagrangian density of fields that can be localized either at the UV or the IR brane. 
The determinant of the induced 4D metric, $g_i^{(4)} =  g^{(5)}/g^{(5)}_{55}$, is just $g^{(5)}(x,y)$ computed at the brane locations, since $g^{(5)}_{55} = -1$.

As stressed above, in the two-brane RS model all fundamental scales are ${\cal O} (\bar{M}_{\rm P})$, whereas relevant energy scales of fields localized at the IR brane are exponentially suppressed (``warped") with respect to the Planck mass. In particular, a bare Higgs mass $m_0$ of the order of the Planck scale becomes exponentially red-shifted at the IR brane, {\em i.e.}, the physical Higgs mass can be of the order of the electroweak scale due to the exponential warping. This is a consequence of the induced metric $ \sqrt{-g_i^{(4)}}$ computed at $y = \pi r_c$, that gives (after rescaling of the Higgs field in order to get a canonically normalised kinetic term)  $m_H = m_0 \times \exp{\left ( - \pi k r_c \right ) }$, being $m_H$ the observed Higgs mass. For $k r_c\sim\mathcal{O}(10)$, one obtains $m_H \sim 100$ GeV, thus solving the SM hierarchy problem. 

The values of $s_i$,
\begin{equation}
s_{\rm UV}= - s_{\rm IR} =  6 \,M_5^3 \,k \,,
\label{eq:branetensionfinetuningRS}
\end{equation}
are chosen to properly glue the metric in the intervals $y \in ]-\pi r_c, 0[$ and $]0, \pi r_c[$, while at the same time 
enforcing reflectivity at the orbifold fixed points $y = 0$ and $y = \pi r_c$. Once gravity is linearised,
\begin{equation}
g^{(5)}_{MN} = \bar g^{(5)}_{MN} + \frac{1}{M_5^{3/2}} \, h^{(5)}_{MN} (x,y) + \dots \, ,
\end{equation}
 the 5D graviton field $h^{(5)}_{MN}(x,y)$ can be decomposed into a KK tower of 4D fields as 
\begin{equation}
    	h^{(5)}_{MN}(x,y) = \sum_n h_{MN}^{(n)}(x) \chi^{(n)}(y) \, ,
\end{equation}
from which we see that the wave-function $\chi^{(n)}(y)$ has dimension $[\chi^{(n)}(y)] = 1/2$.
The 5D graviton is indeed split into three different KK towers of 4D fields: a tower of massive spin-2 excitations, $h_{\mu\nu}^{n} (x)$, called the {\em KK gravitons}; 
a tower of massive spin-1 excitations, $h_{\mu 5}^n (x)$, the {\em KK graviphotons};  and, a tower of scalar excitations, $h_{55}^n (x)$, the {\em KK graviscalars}.
In the so-called {\em unitary gauge}, defined as the gauge where only physical fields propagate, it can be shown \cite{Giudice:1998ck} that the KK tower of massive graviscalars is absent, as they are ``eaten" by the KK tower of graviphotons to get a mass. However, the KK tower of the thus massive graviphotons is also absent, as  in turn they are ``eaten" by the KK gravitons to get their masses. Therefore, the only physical dofs present in the spectrum are the massless modes $h_{\mu\nu}^0$, $h_{\mu 5}^0$ and $h_{55}^0$, plus a tower of (5 dofs) massive KK gravitons. It can also be shown that the massless graviphoton does not couple with matter particles~\cite{Giudice:1998ck}. The eigenfunctions in the extra dimensions are obtained solving the equation of motion
\begin{equation}
\left \{ \partial_{y}^2 - 4 k \partial_{y} + m_{n}^2 e^{2 k y} \right \} \chi^{(n)} (y) = 0
\end{equation} 
in the interval $y \in [0, \pi r_c]$. 
Choosing a conformal coordinate $z$ appropriately \cite{Csaki:2004ay}, the metric can be written as 
$ds^2 = 1/g^2(z) \left ( \eta_{\mu\nu} dx^\mu dx^\nu - dz^2\right)$, 
where
\begin{equation}
    z = \frac{1}{k} \left ( e^{k y} - 1 \right )\,, \; \qquad {\rm for\,\,} \;  y \in [0, \pi r_c] \, ,
\end{equation}
and the ``conformal weight" is
\begin{equation}
    g(z) = k z + 1 \, .
\end{equation}
In terms of the conformal coordinate $z$, the KK graviton wave-functions are:
\begin{equation}
\label{eq:eigenfunctions2branes}
\left \{
\begin{array}{l}
 \hat \chi^{(0)} (z) = \frac{A_0}{\left [ g(z) \right ]^{3/2}} \, , \qquad m_0 = 0 \, , \\
 \\
    \hat \chi^{(n)} (z) =  \sqrt{\frac{g(z)}{k}}\, \left [ 
    A_{1n} \, Y_2 \left ( \frac{m_n}{k} \, g(z)\right ) + 
    A_{2n} \, J_2 \left ( \frac{m_n}{k} \, g(z)\right ) 
    \right ] \, ,
    \end{array}
    \right .
\end{equation}
where $\hat \chi^{(n)}(z) = \sqrt{g(z)} \chi^{(n)}(z)$, and $J_2$ and $Y_2$ are the Bessel functions of the first and the second kind, respectively. 

The non-vanishing KK graviton masses, $m_n$, and the coefficients $A_{0,1n,2n}$ can be found by applying orbifold boundary conditions (BCs), $\chi^{(n)} (y) = \chi^{(n)} (y + 2 \pi r_c)$ and 
$\chi^{(n)}(y) = \chi^{(n)} (-y)$, continuity of the wave-function at the location of the two branes, $\chi^{(n)}(y \to 0^-, \pi r_c) = \chi^{(n)}(y \to 0^+, -\pi r_c)$, and implementing the discontinuity \cite{Shiromizu:1999wj} of the first derivative\footnote{Notice that an alternative way to get the same result proceeds in two steps: first, we implement the BCs relative to the discontinuity of the first derivative of the background metric due to the brane tension terms (at least) localized at $y = 0, \pi r_c$. Then, we solve the Einstein equation for the tensor fluctuation $h_{\mu\nu}$ of the metric, imposing continuity of the wave-function of the KK gravitons and of its first derivative at the brane locations (see Ref.~\cite{Csaki:2000zn} for the details of this procedure).}
 at the location of the two branes (that allows to glue together the background metric on the left and on the right of each brane). One obtains:
\begin{equation}
\label{eq:BCtwobranes}
\left \{
\begin{array}{l}
Y_1 \left ( \frac{m_n}{k}\right ) + \frac{ A_{2n} }{ A_{1n} } \,  J_1 \left ( \frac{m_n}{k} \right ) = 0 \, ,\\
\\
Y_1 \left ( \frac{m_n}{k} \, e^{\pi k r_c} \right ) +  \frac{A_{2n}}{A_{1n}} \, J_1 \left ( \frac{m_n}{k} \, e^{\pi k r_c }\right ) = 0 \, .
\end{array}
\right .
\end{equation}
We define the warping factor as $\omega_{\rm 2b} = \exp(- \pi k r_c)$, with the label ``2b" reminding that this warping is specific to the two-brane setup,
and the mass $m_n$ as $m_n = k \, \left (x_n + \epsilon_n \right ) \, \omega_{\rm 2b} $, where $x_n$ are the zeroes of
the Bessel function of the first kind $J_1 (x)$, thus $m_n \ll k$. Assuming that the shifts $\epsilon_n$ are small quantities, from the first equation in Eq.~\eqref{eq:BCtwobranes} 
we get at leading order in $\omega_{\rm 2b}$
\begin{equation}
\label{eq:twobranesJ2coeff}
A_{2n} = \frac{4}{\pi} \, \frac{1}{x_n^2 \, \omega_{\rm 2b}^2} \, A_{1n} + \dots \, ,
\end{equation}
from which we immediately see that $A_{2n} \gg A_{1n} $, and substituting in the second equation we find
\begin{equation}
\label{eq:twobranesmassshift}
\epsilon_n = \frac{\pi}{4} \, x_n^2 \, \frac{Y_1(x_n)}{J_2(x_n)} \, \omega_{\rm 2b}^2 + \dots \, .
\end{equation}
This means that the mass spectrum is given  by \cite{Davoudiasl:1999jd}
\begin{equation}
\label{eq:twobranesmaspectrum}
m_n = k \, x_n \, \omega_{\rm 2b} + {\cal O}(\omega_{\rm 2b}^3) \, , 
\end{equation}
{\em i.e.}, at leading order the masses are proportional to the zeroes of the $J_1$ Bessel function and are ${\cal O} (\omega_{\rm 2b} M_{\rm P})$. 
The normalisation factors $A_0$ and $A_{1n}$ can be obtained using the orthonormalisation condition,
\begin{equation}
\label{eq:twobranesnormalizationcond}
    2 \int_0^{\bar z}  dz  \hat \chi^{(m)} (z) \, \hat \chi^{(m)} (z)  = 1 \qquad m = 0, 1, \dots \, .
\end{equation}
For the zero mode, we get
\begin{equation}
\label{eq:twobranesnormalizationfactor0}
A_0 = \sqrt{\frac{k}{1-\omega_{\rm 2b}^2}} \, ,
\end{equation}
whereas for the $n$-th mode, we obtain 
\begin{eqnarray}
1 &=& 2 \int_0^{z(\pi r_c)} \, dz \, \frac{g(z)}{k} \,
\left [ A_{1n} \, Y_2 \left ( \frac{m_n}{k} \, g(z) \right ) + A_{2n} \, J_2 \left ( \frac{m_n}{k} \, g(z) \right ) \right ]^2 
    \nonumber \\
    &\simeq& \frac{32}{\pi^2} \, \frac{1}{x_n^6 \, \omega_{\rm 2b}^6} \, \frac{A_{2n}^2}{k^2} \, \int_{x_n \omega_{\rm 2b}}^{x_n} \, du \, u \, J^2_2(u) \, ,
\end{eqnarray} 
where $u = x_n \omega_{\rm 2b} g(z)$. Using standard Bessel function integration tables (see, {\em e.g.}, Ref.~\cite{BesselIntegrals}),
\begin{equation}
\label{eq:twobranescoefficientA2n}
A_{2n} = k \, \frac{1}{J_2(x_n)} \, \omega_{\rm 2b} + \dots \,,
\end{equation}
from which
\begin{equation}
\label{eq:twobranesnormalizationfactorn}
A_{1n} = \frac{\pi}{4} \, k \, \frac{x_n^2}{J_2(x_n)} \, \omega_{\rm 2b}^3 + \dots \, ,
\end{equation}

Therefore, at leading order in $\omega_{\rm 2b}$, the wave-function for the $n$-th KK graviton in terms of the conformal coordinate $z$ is given by
\begin{eqnarray}
\label{eq:twobranefinaleigenfunction}
    \hat \chi^{(n)}(z) &=& 
    \sqrt{k \, g(z)} \, 
    \left [ \frac{\pi}{4} \, \frac{x_n^2}{J_2 (x_n)} \, \omega_{\rm 2b}^3 \,
    Y_2 \left (\frac{m_n}{k} g(z) \right ) -
     \frac{1}{J_2 (x_n)} \, \omega_{\rm 2b} \, 
     J_2 \left (\frac{m_n}{k} g(z) \right )\right ] \\
     && \nonumber \\
    & \simeq &
    \sqrt{k \, g(z)} \, 
    \left [ -
     \frac{1}{J_2 (x_n)} \, \omega_{\rm 2b} \, 
     J_2 \left (\frac{m_n}{k} g(z) \right ) + \dots \right ] \, .
     \nonumber
\end{eqnarray}

For small $n$, the KK graviton modes are far apart from each other and their separation is not uniform,
as it is proportional to the separation of the zeroes of the Bessel function. They are usually treated as independent resonances for LHC searches.
On the other hand, for large $n$, the separation between modes becomes approximately constant. 

It can be shown that the coupling of KK gravitons with fields localized at the IR brane (such as SM fields in the standard RS setup) are universal,
\begin{equation}
{\cal L} = -\frac{1}{\bar M_{\rm P}} \, T^{\mu \nu}(x) h^{0}_{\mu \nu}(x) -\frac{1}{\Lambda_{\rm IR, 2b}} \, \sum_{n=1} T^{\mu \nu}(x) h^{n}_{\mu \nu}(x) \, ,
\end{equation}
and proportional to a single scale,
\begin{equation}\label{eq:2bLambdaIR}
\Lambda_{\rm IR, 2b} = \omega_{\rm 2b} \, \bar M_{\rm P} \, .
\end{equation}
This is not the case for fields localized
in the UV brane, where the coupling of KK gravitons with UV-localized  fields depends on the KK number $n$. In this case, in the two-brane model the coupling of the $n$-th KK graviton with a 4D field at $y = 0$ is proportional to 
\begin{equation}
\label{eq:twobranesKKcouplingUV}
\Lambda^n_{\rm UV, 2b} = c_n \, \omega_{\rm 2b}^{-1} \, \bar M_{\rm P} \, ; \qquad \qquad c_n = \frac{J_2(x_n)}{J_2( \omega_{\rm 2b} x_n)} \sim 
8 \, \frac{J_2(x_n)}{x_n^2\,\omega_{\rm 2b}^{2}} + {\cal O}(\omega_{\rm 2b}^2) \, ,
\end{equation}
where the last expression holds as long as $\omega_{\rm 2b} x_n \ll 1$.

In the absence of a stabilising mechanism for the distance between the branes, the graviscalar dof would be massless. 
A possible interpretation for that is the fact that the RS solution is obtained for any choice of $r_c$. Therefore, choosing a specific value for the brane separations corresponds to a spontaneous breaking of translational invariance along the extra dimension, with a resulting massless Goldstone boson. To generate a mass for it, one needs to introduce an explicit breaking source for translational invariance. The compactification radius $r_c$ may indeed be fixed to a specific value by means of an additional bulk field $\varphi$ with both a (trivial) bulk potential $V_{\rm bulk}(\varphi) = m^2 \varphi^2$ and two localized potentials $V_i(\varphi)$ chosen appropriately. This ``stabilisation" mechanism, that fixes dynamically the value of $r_c$, is known as {\em Goldberger-Wise mechanism}, and it was suggested in Refs.~\cite{GW1,GW2}. After integration over the extra dimension of the bulk field action, 
a potential for the graviscalar is found, whose minimum fixes the relation between $k r_c$ and the ratio of the localized potentials $V_i$ at $y = 0$ and $y = \pi r_c$. Eventually, 
expanding over this minimum, a mass term for the  graviscalar field is generated. It can be shown that the mass of this state (by now called the ``radion") is suppressed by the backreaction of the metric compared to the masses of the KK gravitons. As the backreaction depends on the Goldberger-Wise potential terms, we can treat the mass of the radion, $m_r$, as a free parameter. On the other hand, the mass of the zero mode of the bulk scalar field is heavy $\sim\mathcal{O}(m)$.
The KK tower of the bulk scalar $\varphi$, $\varphi_{(n)}$, would also be part of the physical spectrum. 
However, it can be shown that their interactions with other brane fields vanish in the limit of no backreaction of the bulk field over the background metric
 (see Ref.~\cite{Csaki:2000zn}). Therefore, they also decouple and will not be considered when studying the phenomenology of the model. 

In the ``standard" RS setup, 
in addition to the SM fields, new matter fields may be located at the IR-brane. If stable or sufficiently long lived, these fields may act as DM, as they interact gravitationally with SM fields through KK gravitons, thus feeling an enhanced gravity. Attempts in this direction to explain the observed DM relic abundance, $\Omega_{\rm DM}$, have been presented in Refs.~\cite{Lee:2013bua,Rueter:2017nbk,Carrillo-Monteverde:2018phy,Rizzo:2018joy,Folgado:2019sgz,Folgado:2020vjb}. The idea is that extra-dimensionally-enhanced gravitational interactions may be large enough to reproduce $\Omega_{\rm DM}$ within the freeze-out paradigm, either through ${\rm DM \, DM} \to { \rm SM \, SM}$ or ${\rm DM \, DM } \to G_m \, G_n, G_m \, r, r \, r$ (with $G_n$ a KK graviton). However, in all of these references diagrams involving gravitational triple vertices (such as $G_k G_m G_n$, $G_k G_m r$, $G_k r r$ and $r r r$) have been overlooked. 
Due to this omission, the cross-section ${\rm DM \, DM } \to G_m \, G_n$ violates unitarity as ${\cal O}(s^3)$, thus rapidly becoming the dominant channel for large enough $s$. 
Even though ${\cal O}(s)$ unitarity violation is to be expected (as linearised gravity is a non-renormalizable effective theory), the divergence with $s$ is indeed too large. 

It can be shown that the DM relic abundance can be 
obtained mainly via DM annihilations into SM particles. For this reason,
LHC null results for resonance searches \cite{Aaboud:2017yyg,ATLAS:2017wce,CMS:2018thv} constrain significantly the parameter space for this option. 
After taking into account all the experimental constraints, the allowed region in the parameter space for which 
a massive scalar gravitationally interacting with SM particles in a 5D RS setup can lead to the observed DM relic abundance via freeze-out 
is very small indeed: the DM mass $m_{\rm DM}$ should be ${\cal O} (10)$ TeV, 
the first KK graviton mass larger than 5 TeV and the effective scale $\Lambda_{\rm IR, 2b}$ in the range $\Lambda_{\rm IR, 2b} \in [5,10]$ TeV 
(see Erratum of Ref.~\cite{Folgado:2019sgz}).

In order to lessen the impact of LHC data on the allowed parameter space of a model with DM in the extra dimension,  in the following we consider the possibility to split the brane that may solve the hierarchy problem (the IR brane), from a second one in which DM particles live (the deep IR, or DIR brane). In this way, we gain substantial freedom that allows us to address both problems simultaneously. 

\subsection{A three-brane Randall-Sundrum scenario} \label{sect:threebranes}

A three-brane model was first presented long ago, in Ref.~\cite{Lykken:1999nb}, with the goal of merging the virtues of the RS model discussed above
\cite{RandallLisa1999LMHf} (also called RS1) with the so-called RS2 model \cite{RandallLisa1999AAtC} (that differs from the first in that the second brane is moved to infinity). 
The phenomenology of the two two-brane models is quite different: whereas the first model is designed to address the SM hierarchy problem by locating SM fields at the $y = \pi r_c$ fixed point of the orbifold (thus achieving that energy scales as seen as from the IR-brane point of view are warped down to the electroweak scale, even though they are 
${\cal O}(M_{\rm P})$ at the fundamental level), the second model aims at showing that 4D-gravity can be recovered in a 5D space-time if the curvature in the extra-dimension is 
``large enough" to prevent
low-energy excitations of the graviton to enter it. 

Once three branes are considered, several options arise, though. In order to reproduce a background metric valid in the whole space-time, different configurations are possible depending on the sign of the brane tension terms $s_i$. Two options are typically considered: the ``$+ - +$" option~\cite{Kogan:1999wc} 
(in which first and third branes have a positive tension, whereas the tension of the middle one is negative) and the ``$+ + -$" option~\cite{Lykken:1999nb}. 
In the former case,  it was shown that the first KK mode of the graviton is extremely light. This mode, together with the graviton zero mode, gives rise to an effective 4D bi-gravity theory \cite{Kogan:2000cv}. At the same time, it accomplishes the goal of extending the RS2 model so as to address the hierarchy problem (taking advantage of the intermediate brane).
Albeit with interesting phenomenological consequences, we are interested here in the latter three-brane scenario,  the ``$+ + -$" configuration. This model is better suited to phenomenology, as it allows us to play with the location of different branes, and thus, achieving different warpings of the energy scales we are interested in.  A $Z_2$ orbifold symmetry with compactification radius $r_c$ is also considered, as in the RS two-brane setup. Two branes are still located at the orbifold fixed points, $y = 0$ and $y =  L_2 = \pi r_c $, 
whereas the third brane is located at an arbitrary point in between, $y = L_1$. In the two bulk sub-regions,  $y \in ]0, L_1[$ and $y \in ]L_1, L_2[$, two different 5D cosmological 
constants are considered, $\Lambda_1$ and $\Lambda_2$. In order to get a stable background metric, the tensions in the three branes must be related to the two cosmological constants and to their difference.

The action of the model is given by
\begin{eqnarray}\label{eq:threebranesaction}
{\cal S} &=&  {\cal S}_{\rm grav} + {\cal S}_{\rm branes} = - M_5^3 \, \int d^4 x \int_0^{L_2} dy \sqrt{g^{(5)}} \, 
\left \{ R^{(5)} + 2 \Lambda (y) \right \} 
\nonumber \\
&+& \int d^4 x \int_0^{L_2} \sqrt{-g_{\rm UV}} \, \delta (y) \left [ -s_{\rm UV} + \dots \right ] \nonumber \\
&+& \int d^4 x \int_0^{L_2} \sqrt{-g_{\rm IR}} \, \delta (y - L_1) \left [ -s_{\rm IR} + {\cal L}_{\rm SM} \right ] \nonumber \\
&+& \int d^4 x \int_0^{L_2} \sqrt{-g_{\rm DIR}} \, \delta (y - L_2) \left [ -s_{\rm DIR} + {\cal L}_{\rm DM} \right ] \,,
\end{eqnarray}
where 
$y_{\rm UV} = 0, y_{\rm IR} = L_1$ and $y_{\rm DIR} = L_2$ are the branes' locations, 
$\Lambda (y) = \Lambda_1$ for $y \in ]0, L_1[$, $\Lambda (y) = \Lambda_2$ for $y \in ]L_1, L_2[$ is the cosmological constant in the two bulk sub-regions; $g_{MN}^{(5)}$ is the 5D metric, with determinant $g^{(5)}$, whereas $g_i$ are the determinants of the induced metric on the three branes, $g_i (x) = g^{(5)}(x,y_i)/g_{55}^{(5)}$; eventually, $M_5$ is the fundamental 5D gravitational scale. Notice that, being the DM and SM particles on different branes, the Higgs portal is not allowed.

The two curvatures, $k_1, k_2$, are related to the two cosmological constants as
\begin{equation}
k_1 = \sqrt{\frac{-\Lambda_1}{6 M_5^3}} \, ; \qquad k_2 = \sqrt{\frac{- \Lambda_2}{6 M_5^3}} \, .
\end{equation}
The brane tensions $s_i$ must be chosen appropriately in order to glue the background metric piecewise as:
\begin{equation}
    s_{\rm UV} = 6 M_5^3 k_1 \, ; \qquad 
    s_{\rm IR} = 3 M_5^3 (k_2 - k_1)\, ; 
    \qquad s_{\rm DIR} = - 6 M_5^3 k_2 \, ,
\end{equation}
where $k_2 > k_1$ in order to enforce the ``+ + -" brane configuration. Notice that, in the limit\footnote{We stress here that the limit of vanishing intermediate brane ($k_2 \to k_1$)
is formally and phenomenologically very different from the limits in which the intermediate brane moves either to its left, $L_1 \to 0$, or to its right, $L_1 \to L_2$.}
 $k_2 \to k_1$ we recover the standard RS BCs (as the intermediate brane ``vanishes").
 
In our case, the DM is located at the DIR brane, whereas the SM is at the IR brane.

In Ref.~\cite{Kogan:2000xc} it was shown that the reduced Planck mass of the ``$+ + -$" model is related to the curvatures in the bulk sub-regions, $k_1$ and $k_2$, and to the lengths of the two segments, $L_1$ and $L_2 = L_1 + \Delta L$, as 
\begin{equation}
\label{eq:Mplanck3branes}
    \bar M_{\rm P}^2 = M_5^3 \left [ 
    \frac{1}{k_1} \left ( 1 - e^{-2 k_1 L_1}\right )  
    + \frac{1}{k_2} e^{- 2 k_1 L_1} 
    \left ( 1 - e^{-2 k_2 \Delta L}\right ) 
    \right ] \, .
\end{equation}
This expression reduces to the standard two-brane RS relation both for $\Delta L \to 0$ and $L_1 \to 0$ (for which $\Delta L \to L_2$). These two limits are not the only two cases for which the two-brane setup is recovered, though. For large enough $k_1 L_1$ and $k_2 L_2$, the four dimensionful quantities $M_5, k_1, k_2$ and $\bar M_{\rm P}$ can be taken to be of the same order. It can be shown that $k_2$ must be $k_2 \gtrsim k_1$ in order to avoid tachyonic modes, $k_2 = k_1 + \delta k$. Considering $M_5 \gtrsim k_2 \gtrsim  k_1$ (so that the fundamental Planck scale $M_5$ is the largest scale in the framework), we
can see that in the limit $\delta k/k_1 \ll 1$ we again
obtain Eq.~\eqref{eq:Mplanck3branes}, {\em i.e.} the two-brane setup relation. Since the brane tension of the
intermediate brane is proportional to $\delta k/k_1$, 
and it could thus be much smaller than that of the UV  
and DIR branes, we call this limit 
{\em evanescent brane}. For simplicity, in the rest of
the paper we remove the subindex and define $k_1 \equiv k$.

In order to derive the coupling among the 4D fields (SM and DM) and the bulk fields (radion and KK gravitons) we need to compute the wave-functions of the latter. Let us first consider the graviton KK expansion (we only focus on the spin-2 resonances), 
\begin{equation}\label{eq:KKmodes}
    h_{\mu\nu}^{(5)}(x,y)=\sum_{n}h_{\mu\nu}^n(x)\chi^{(n)}(y)\,.
\end{equation}
Now we consider the limit $\delta k\ll k$, in which the KK graviton wave-functions formally coincide with those of the two-brane setup. It is convenient to express them in terms of the physical coordinate $y$ as
\begin{equation}
    \chi^{(0)}=\sqrt{\frac{k}{1-e^{-2kL_2}}}\simeq \sqrt{k}\,, \qquad \chi^{(n\neq0)}(y)\simeq\sqrt{k}\,{e^{k(2y-L_2)}}\frac{J_2\left(x_n e^{k(y-L_2)}\right)}{J_2(x_n)}\,,
\end{equation}
where $J_1(x_n)=0$. The KK graviton masses are obtained imposing the BC's similarly to the two-brane case. We find
$m_n \simeq k \, x_n \, \xi\,\omega$, where we have defined 
\begin{equation}
    \xi\equiv \exp[-k(L_2-L_1)]\,,\qquad \omega=\exp(-kL_1)\,.
\end{equation}
The 5D graviton couples to matter localized at one of the two branes as
\begin{equation}
   \mathcal{L}_h =\sum_{i={\rm IR},{\rm DIR}}\frac{1}{M_5^{3/2}}\,h_{\mu\nu}(x,y_i)\,T^{\mu\nu}_i(x)\,,
\end{equation}
being $T_i^{\mu\nu}$ the energy momentum tensor of matter
at $y = y_i$.
Using the KK expansion of Eq.~\eqref{eq:KKmodes} we arrive at 
\begin{equation} \label{eq:gEFT}
	{\cal L}_{h} = \frac{1}{\bar M_{\rm P}} \, h_{\mu\nu}^{(0)} (x) \, T^{\mu\nu} (x) + 
	\sum_{n = 1}^\infty \frac{1}{\Lambda^n_{\rm IR}} \, h_{\mu\nu}^{(n)} (x) \, T^{\mu\nu}_{\rm IR} (x) \, + 
	\sum_{n =1}^\infty \frac{1}{\Lambda_{\rm DIR}} \, h_{\mu\nu}^{(n)} (x) \, T^{\mu\nu}_{\rm DIR} (x) \, ,
\end{equation}
where KK gravitons (with $n\neq0)$ couple to SM matter localized at the IR brane with effective inverse coupling
\begin{equation}\label{eq:LambdaIR}
\Lambda_{\rm IR}^n = \frac{\omega}{\xi}\bar{M}_P\frac{J_2(x_n)}{J_2(\xi x_n)} \, \sim 8 \,  \frac{\omega \bar M_{\rm P}}{\xi^3}\frac{J_2(x_n)}{x_n^2}  + {\cal O}(\xi^2)\, ,
\end{equation}
where the expansion in Eq.~\eqref{eq:LambdaIR} is only valid if $x_n\,\xi\ll1$.
Notice that the coupling depends on the KK number $n$, so that every mode interacts with a different strength.

On the other hand, all KK gravitons couple to DM, localized at the DIR brane, with the same effective inverse coupling,
\begin{equation}\label{eq:LambdaDIR}
    \Lambda_{\rm DIR}=\omega\,\xi \,\bar{M}_P\,.
\end{equation}
Finally, the massless graviton $h_{\mu\nu}^{(0)}$ couples universally with all matter fields, with Planck-suppressed coupling.

The expressions obtained in the limit $\delta k/k\ll 1$ can be generalized using the full three-brane KK graviton wave-functions with $k_1\neq k_2$. The explicit computation can be seen in the companion paper, Ref.~\cite{Donini:2025qrf}, in which some of us discuss the formal aspects of our three-brane setup. Here, we summarize our main findings: fields on the DIR brane couple universally to all KK modes, proportionally to $1/\Lambda_{\rm DIR}$, formally identical to Eq.~\eqref{eq:LambdaDIR} with the replacements $\xi\to\exp[-k_2(L_2-L_1)]$ and $\omega\to \exp(-k_1L_1)$. Fields in the IR brane couple to the $n$-th KK graviton with coupling $1/\Lambda^n_{\rm IR}$ multiplied by the factor $(k_1/k_2)^{3/2}$.

Eventually, a dependence of the KK graviton masses on the ratio between the two curvatures, $(k_1/k_2)$, arises when
corrections to the leading terms are considered. 
It can be shown that, for
\begin{equation}
m_n = k_2 \left (x_n + \epsilon_n\right) \, \xi \, \omega
\end{equation}
we have 
\begin{equation}
\epsilon_n = - \frac{\pi}{4} \, 
\left ( \frac{k_1}{k_2}\right ) \, \xi^2 \, \omega^2 
\, \left [ 
x_n^2 \, \frac{Y_1(x_n)}{J_2(x_n)} 
\right ] + {\cal O} (\omega^4) \, .
\end{equation}
As it was the case for the two-brane scenario, in the absence
of a stabilising mechanism the graviscalar dof would be massless. An explicit breaking of translational invariance in the extra dimension that would give a mass
to the graviscalar is obtained by adding a scalar bulk field
$\varphi$, with trivial (but possibly different) potential in each segment, $V_{\rm bulk, i} = m_i^2 \varphi^2$ (with $i = 1,2$ referring to $y \in [0,L_1]$ and $y \in [L_1,L_2]$, respectively). In Ref.~\cite{Donini:2025qrf} we explicitly show
how, introducing a suitable metric for each segment, it is possible to reproduce the {\em Goldberger-Wise mechanism} for
our three-brane setup, obtaining both a ``light" and a ``heavy" radion. The light dof has a mass proportional to $(m_2/k_2)^2$ 
as in the two-brane setup. The ``heavy"
radion decouples from the low-energy spectrum in the limit $\delta k\ll k$, as its mass
is proportional to $1/(k_2 - k_1)=1/\delta k $. Eventually, the two would-be
zero modes of the bulk field $\varphi$ have a mass proportional 
to $m_i$ and, thus, also decouple from the low-energy spectrum. 
In summary, also in the three-brane setup one can consider only one light degree of freedom, the (light) radion $r$. The couplings of the radion with matter are derived in a similar way to the ones of the gravitons. In~\cite{Donini:2025qrf} we perform the full computation, while in App.~\ref{app:Int} we comment on the limit $\delta k\ll k$. We find 
\begin{equation} \label{eq:radEFT}
\mathcal{L}_r=\frac{1}{\sqrt{6}\Lambda_{\rm IR}}rT_{\rm IR}+\frac{1}{\sqrt{6}\Lambda_{\rm DIR}}rT_{\rm DIR}+\frac{1}{8\pi\sqrt{6}\Lambda_{\rm IR}}r\left(\alpha_{\rm em}C_{\rm em}F_{\mu\nu}F^{\mu\nu}+\alpha_s C_3 G_{\mu\nu}^aG_a^{\mu\nu}\right),
	\end{equation}	
where $T=T_\mu^\mu$ is the trace of the energy-momentum tensor, while the last two terms represent interactions with photons
and gluons and are proportional to the corresponding trace anomalies. Radion interactions with matter at the DIR brane are again controlled by the inverse coupling $\Lambda_{\rm DIR}$ given in Eq.~\eqref{eq:LambdaDIR}, as for gravitons. On the contrary, radion interactions with SM particles at the IR brane are controlled by the inverse coupling
\begin{equation}\label{eq:IRscale}
    \Lambda_{\rm IR} \equiv  \frac{\omega}{\xi} \, \bar M_{\rm P}\, .
\end{equation}
The couplings of the KK gravitons and the radion are summarized in Table~\ref{table:interactions}. Radion interactions with the SM (IR brane) and DM (DIR brane) are proportional to $1/\Lambda_{\rm IR}$ and $1/\Lambda_{\rm DIR}$, respectively, whereas the couplings of the KK gravitons with the SM and DM are proportional to $1/\Lambda^n_{\rm IR}$ and $1/\Lambda_{\rm DIR}$, respectively. Out of the limit $\delta k/k \ll 1$, 
it can be shown that the coupling of fields on the
IR brane with the radion is $1/\Lambda_{\rm IR}$ multiplied by the factor $(k_1/k_2)^{1/2}$.

\begin{figure}[htbp]
	\centering
\includegraphics[width=0.5\textwidth]{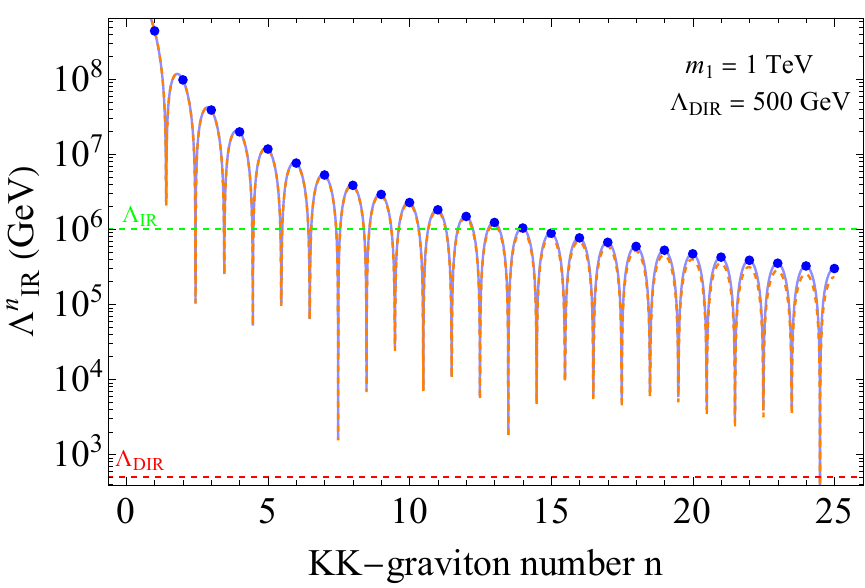}~~\includegraphics[width=0.5\textwidth]{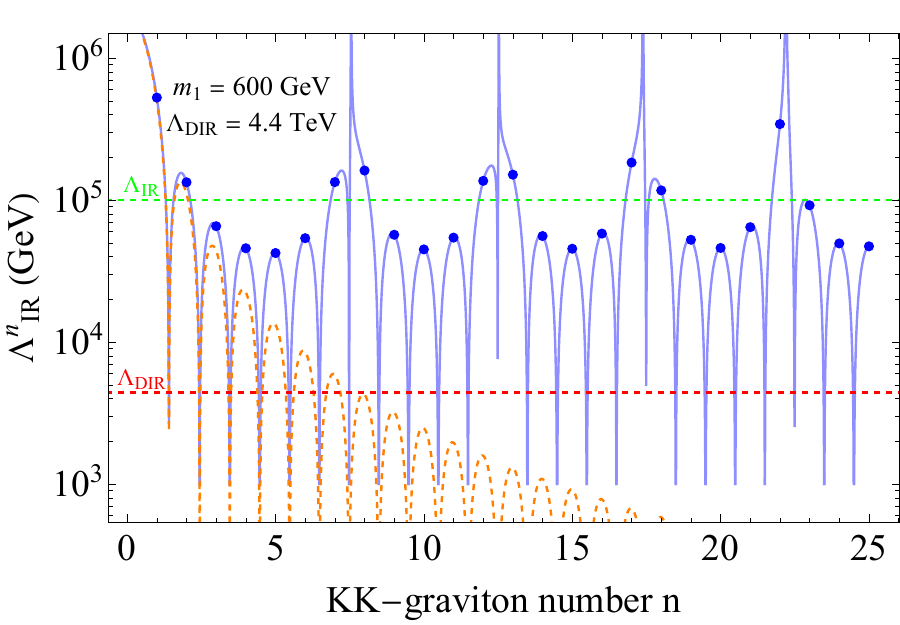}
	\caption{Couplings of the KK gravitons with matter in the IR brane (SM), $\Lambda_{\rm IR}^n$, as a function of the KK mode $n$ for different fixed values of $\Lambda_{\rm DIR}$, $\Lambda_{\rm IR}$ and $m_{1}$. The continuous blue line corresponds to the analytical function for $\Lambda_{\rm IR}^{n}$ in Eq.~\eqref{eq:LambdaIR} and takes  physical values at the discrete blue dots, each one representing a KK mode. The orange dashed curves corresponds to the analytical approximation valid for $\xi x_n\ll1$.} \label{fig:KKcouplingsa}
\end{figure}

In Fig.~\ref{fig:KKcouplingsa} we show the inverse couplings of the first 25 KK gravitons to matter in the IR brane (SM), $\Lambda_{\rm IR}^n$, for fixed values of $\Lambda_{\rm IR}$ and $\Lambda_{\rm DIR}$. If the two scales are very hierarchical, $\Lambda_{\rm IR}\gg \Lambda_{\rm DIR}$, we can expand $\Lambda_{\rm IR}^n$ in $\xi x_n= e^{-k (L_2-L_1)} x_n = \sqrt{\Lambda_{\rm DIR}/\Lambda_{\rm IR}}\, x_n\ll1$ up to large values of the KK number $n$ (see left plot in Fig.~\ref{fig:KKcouplingsa}). Thus, as long as the approximation is valid, the higher modes in the graviton KK tower couple to SM matter with stronger couplings. On the other hand, when $\xi x_n\gtrsim1$, the perturbative expansion breaks down and we must use the exact formula. This is particularly important when the hierarchy between $\Lambda_{\rm IR}$ and $\Lambda_{\rm DIR}$ is not very large (see right plot of Fig.~\ref{fig:KKcouplingsa}). We observe that the inverse couplings of KK gravitons to matter oscillate around the scale $\Lambda_{\rm IR}$ as long as $n$ grows (eventually they all collapse to the constant value $\Lambda_{\rm IR}$ in the limit $\Lambda_{\rm DIR}\to\Lambda_{\rm IR}$).

\begin{table}[t!] 
			$$\begin{array}{c|c|c|c|c}
				\rowcolor[HTML]{C0C0C0} 
				\text{Brane} & \text{Particle} & h_{\mu\nu}^0  & \text{Massive KK } h_{\mu\nu}^{n\geq1}  & \text{Radion }   \\ 
                \hline 
                & & & & \\
				\text{IR}  & \text{SM} & \bar{M_{\rm P}}  &  
				\Lambda_{\rm IR}^n  \equiv \omega\,\xi^{-1}\,\bar{M}_{\rm P}\frac{J_2(x_n)}{J_2(\xi\, x_n)}=\Lambda_{\rm IR}\,\frac{J_2(x_n)}{J_2(x_n\,\sqrt{\Lambda_{\rm DIR}/\Lambda_{\rm IR}})} &
				\sqrt{6}\,\Lambda_{\rm IR} \equiv \sqrt{6}\,\omega\,\xi^{{-1}}\,\bar{M_{\rm P}}   \\
                                & & & & \\
                \text{DIR} & \text{DM}   & \bar{M_{\rm P}}  &
				\Lambda_{\rm DIR} \equiv \omega\,\xi\, \bar{M_{\rm P}}&  
				\sqrt{6}\,\Lambda_{\rm DIR}\equiv\sqrt{6}\,\omega\,\xi \,\bar{M_{\rm P}}\ 
			\end{array}$$
		\caption{Inverse couplings of the SM (IR brane) and DM particles (DIR brane) with massless gravitons, KK gravitons and radions in the third, fourth and fifth columns, respectively, in the limit $\delta k/k \ll 1$ of the three-brane setup.
		Note how KK gravitons and the radion couple with the same strength to DM particles (up to a factor $\sqrt{6}$), while graviton interactions with SM particles depend on the KK number $n$. }
			\label{table:interactions}
		\end{table}	

It is straightforward to show (see App.~\ref{app:Int}) that the bare mass $m_0$ of a matter field localized at a brane at $y=L$ is warped down as $m=\exp(-k L)\,m_0$.
Thus, any matter field located at the IR-brane has a mass that is warped down as
\begin{equation}\label{eq:hierarchy}
m_{\rm IR} =  \sqrt{\Lambda_{\rm IR} \, \Lambda_{\rm DIR}} \, \frac{m_0}{\bar M_{\rm P}} = \omega \,  m_0\, ,
\end{equation}
whereas fields located at the rightmost (DIR) brane have warped masses
\begin{equation}
m_{\rm DIR}  = \Lambda_{\rm DIR} \, \frac{m_0}{\bar M_{\rm P}} = \xi \, \omega \,m_0  \, ,
\end{equation}
This means that, if we place the SM and DM fields on different branes, we have some flexibility to address simultaneously 
the SM hierarchy problem and the DM hierarchy problem (namely, explaining why the experimentally observed Higgs mass is ${\cal O}(\Lambda_{\rm EW})$, 
and not as large as the would-be SM cut-off, and why the (scalar) DM mass could be even lighter than that). The two scales, $\Lambda_{\rm DIR}$ and $\Lambda_{\rm IR}$, will be used as free parameters of the model in our phenomenological analysis.

As a final comment, it is interesting to notice that the identification of a dual picture of a multi-brane setup has been overlooked in the literature. Although it would be interesting  to investigate this topic, this goes beyond the scope of the current paper. Furthermore, to the best of our knowledge, a duality for a multi-brane setup is not guaranteed to exist. We signal that the authors of Ref.~\cite{Lee} refer to a dual interpretation of their setup. However, such duality holds for the effective two-brane setup and not for the fundamental three-brane setup.

\vspace{0.5cm}

\section{Decay rates of the KK gravitons and radions}   \label{sec:decays}

Generally speaking, the radion and the lowest KK gravitons must decay fast enough not to overclose the Universe. Furthermore, if they are lighter than the DM candidate,
their decays into SM particles could inject a significant amount of energy into the SM thermal bath. This is constrained by measurements of the abundances of light elements from Big Bang Nucleosynthesis (BBN) as well as by the observations of anisotropies in the Cosmic Microwave Background (CMB). 
Here,  we conservatively require that these decays take place before the onset of BBN. This translates into a constraint on the radion/graviton lifetimes, namely $\tau_{r,G}\lesssim 1$ s.

\begin{figure}[t]
    \centering
\includegraphics[width=0.8\textwidth]{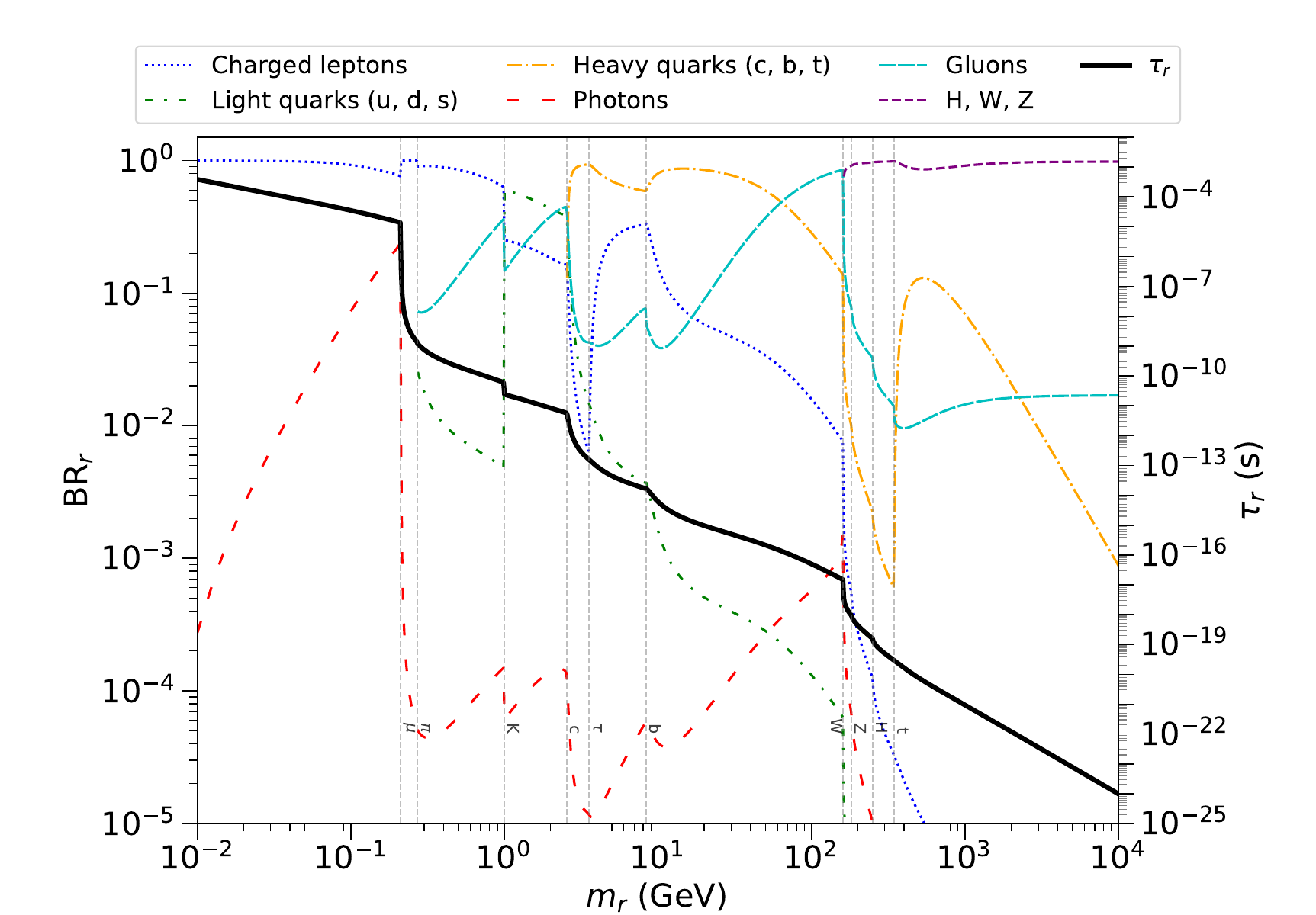}    	
\caption{
Branching ratios of the radion as a function of its mass. The lifetime $\tau$, shown in the right axis, is calculated for $\Lambda_{\rm IR}=100\, \text{TeV}$.
}
\label{fig:decaysr} 
\end{figure}

In App.~\ref{app:decayrates} we provide the decay rates of the radion and the KK gravitons. In most of the parameter space, the KK graviton dominantly decays into radions (being the lightest bulk particles) and/or lighter gravitons (which in turn decay into radions), with $\mathcal{O}(1)$ branching ratios (BRs). Heavy enough KK graviton modes also decay into DM particles with similar BRs. 
On the other hand, the radion decays into SM particles. In Fig.~\ref{fig:decaysr} we plot the BRs (left axis) of the radion into different SM channels (leptons, quarks, gauge bosons) as a function of the radion mass. We also show the total lifetime in seconds (right axis) for $\Lambda_{\rm IR}=100\,\text{TeV}$. The thresholds of different channels are highlighted with gray vertical dashed lines. From now on, we assume that the decay into DM is kinematically closed, i.e., $m_r < 2\, m_{\rm DM}$.

As can be observed, if $m_r\lesssim$ GeV, the dominant decay channels are charged leptons: $e^+e^-$ for $2m_e\lesssim m_r\lesssim 2m_\mu$ and $\mu^+\mu^-$ 
for $2m_\mu\lesssim m_r\lesssim 2m_K$. 
Lighter radions, with $m_r<2m_e$, can only decay into photons with a loop-suppressed decay rate.
We approximate the decay rates into light mesons, when kinematically allowed, by the gluon and light quark BRs, assuming that hadronization will lead to $\mathcal{O}(1)$ corrections~\cite{Bijnens:2001gh}.
For radion masses in the $[1,100]$ GeV range, the dominant decay channels are either heavy quark pairs ($\bar{b}b$, $\bar{c}c$) or light mesons. For heavier masses, radions decay into Higgs and EW gauge bosons. Since we only consider masses such that $m_r>2m_e$, the radion's lifetime is always much shorter than a millisecond, as shown in Fig.~\ref{fig:decaysr}. Thus, we safely evade all cosmological constraints. On the other hand, the decay channels of the radion (and the gravitons) are relevant to constraint the DM parameter space using indirect detection searches, as we discuss in Sec.~\ref{sec:indirect}.

\begin{figure}[!htbp]
    \centering
\includegraphics[width=0.73\textwidth]{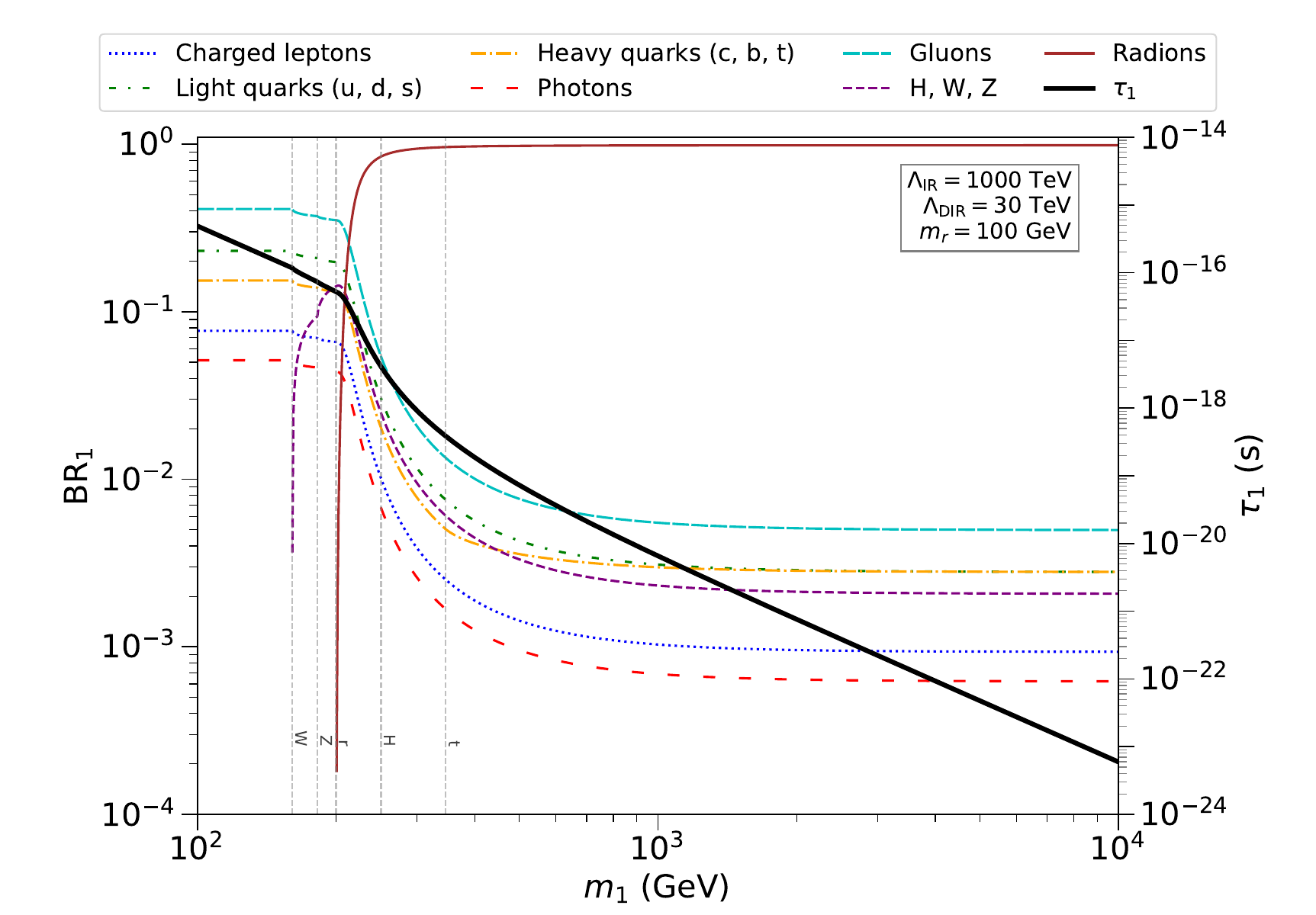}
\includegraphics[width=0.73\textwidth]{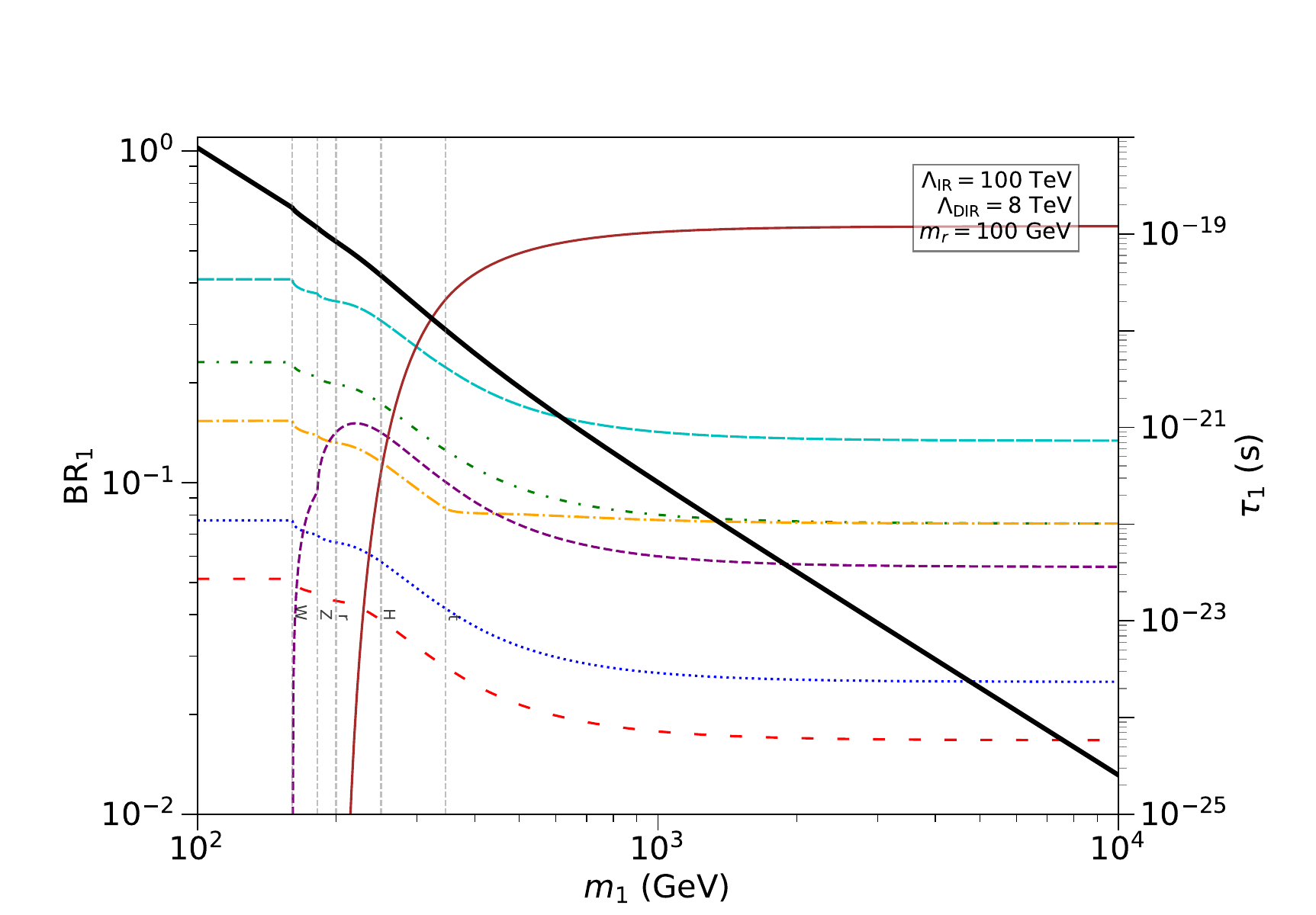}
\includegraphics[width=0.73\textwidth]{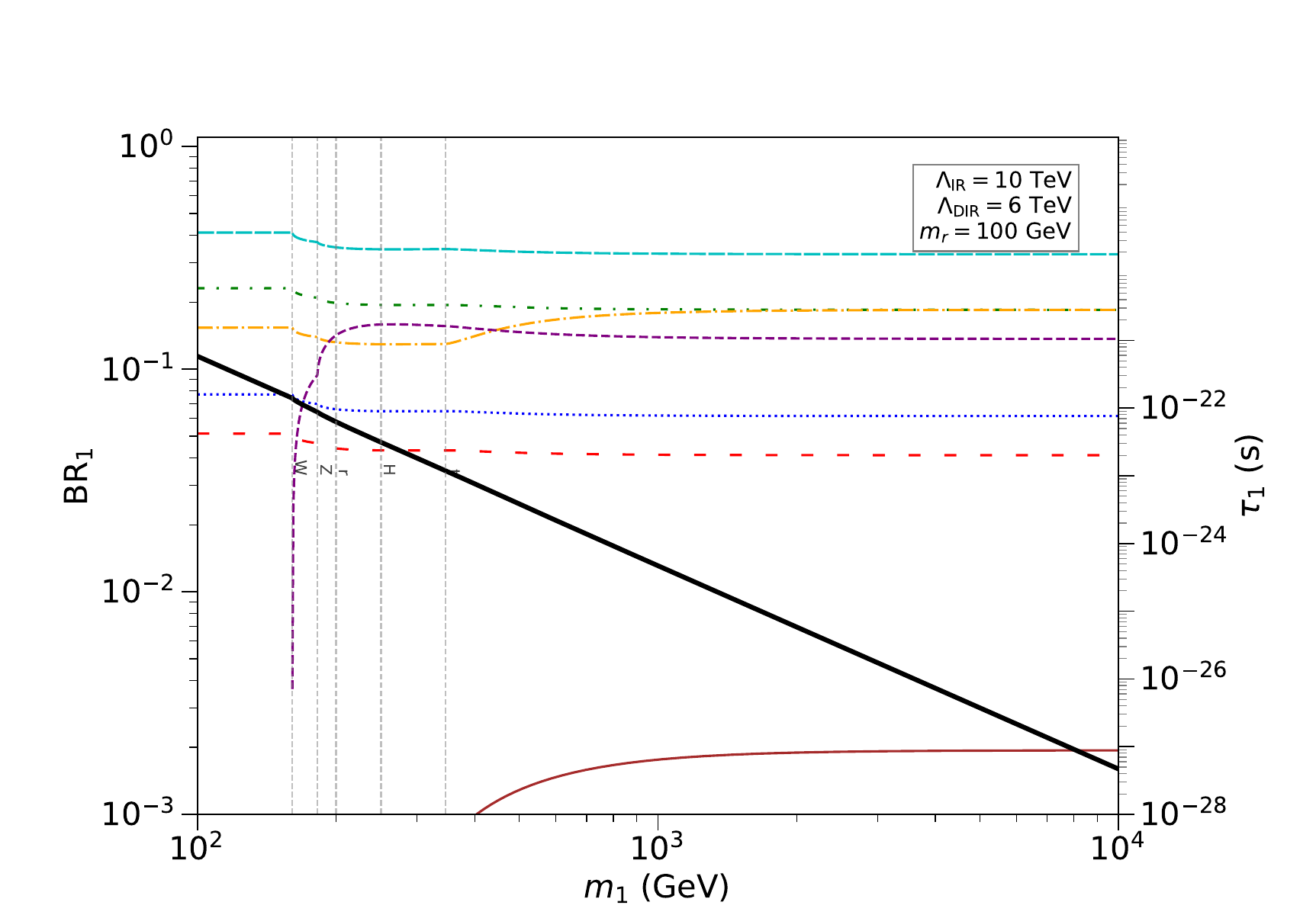}
\\
\caption{Similar to Fig.~\ref{fig:decaysr} for the decays of the first KK graviton for different values of $\Lambda_{\rm IR},\,\Lambda_{\rm DIR}$.}\label{fig:decaysG} 
\end{figure}

In Fig.~\ref{fig:decaysG} we show the BRs (left axes) and the lifetime (right axes) of the first KK graviton for $m_r = 100$ GeV and different values of $\Lambda_{\rm IR}$ and $\Lambda_{\rm DIR}$. We can see that the graviton's lifetime is much shorter than the radion's one. As can be observed, whenever the decay channel of the gravitons into radions is kinematically open, it dominates the decays, unless the two scales $\Lambda_{\rm IR}$ and $\Lambda_{\rm DIR}$ take similar values, $\Lambda_{\rm DIR}\lesssim \Lambda_{\rm IR}$.

	\section{Dark matter relic abundance} \label{sec:relic}

\subsection{Thermal freeze-out}	 

The interactions among DM, bulk particles (KK gravitons and the radion) and SM particles are strong enough to keep the three in thermal equilibrium in the early Universe as long as $\Lambda_{\rm IR} \lesssim 10^8$ GeV.\footnote{If the radion is sufficiently light, the dynamics may be more involved. Indeed, the radion could temporarily decouple from the SM thermal bath before the DM particle and re-enter thermal equilibrium at later times. Ref.~\cite{Koutroulis:2024wjl} showed that this can affect the computation of the relic density for $m_r<$ MeV. We do not consider this possibility in this paper.} For larger values of $\Lambda_{\rm IR}$, DM interactions with SM particles become too weak to establish thermal equilibrium, resulting in a secluded DM scenario. In such a case, if graviton interactions are sufficiently strong to maintain equilibrium among themselves, the DM temperature will differ from that of the SM sector. In the rest of the paper we assume $\Lambda_{\rm IR} \ll 10^8$ GeV so that all the particles are in thermal equilibrium with the same temperature $T$.

In this scenario, the DM number density $n_{\rm DM}$ follows its thermal equilibrium distribution $n_{\rm DM}^{\rm eq}(T)$, being $T$ the SM bath temperature, as long as the DM annihilation rate, $\Gamma_{\rm ann}(T)=n_{\rm DM}(T)\med{\sigma v}$, is larger than the Hubble expansion rate, which in a radiation-dominated universe is given by $H(T)\simeq \sqrt{4\pi^3/45}\,g_*(T)T^2/M_{\rm Pl}$. Here $\med{\sigma v}$ is the thermally-averaged total DM annihilation cross-section (being $v$ the relative velocity of the DM particles), which we will describe more precisely in the next section, and $g_*$ is the number of relativistic degrees of freedom in the thermal bath. When the temperature decreases so much that the annihilation rate falls below the Hubble rate, the DM particle decouples from the thermal bath (\textit{freeze-out}) leaving a constant number density normalised to the entropy density, i.e., the yield $Y_{\rm DM}\equiv n_{\rm DM}/{s}$, with $s=2\pi^2g_s(T)T^3/45$, being $g_s$ the number of relativistic degrees of freedom in entropy present in the thermal bath ($g_s$ and $g_*$ only differ at temperature below $\sim 0.5$ MeV).

More precisely, the evolution of the DM number density is governed by the Boltzmann equation\footnote{A more precise analysis would require the solution of a system of multiple coupled Boltzmann equations for the DM, the radion and the KK gravitons. 
If the radion and the gravitons are in thermal equilibrium with the SM bath during the time relevant for DM freeze-out,
the system reduces to the single Boltzmann equation provided in the main text, Eq.~\eqref{eq.BEQ}.}
    \begin{equation}\label{eq.BEQ}
        \frac{dn_{\rm DM}}{dt}=-3H(T)n_{\rm DM}-\med{\sigma v}\left[n_{\rm DM}^2-(n_{\rm DM}^{\rm eq})^2\right]\,.
    \end{equation}
The solution of Eq.~\eqref{eq.BEQ} provides the DM relic abundance, $\Omega_{\rm DM}=\rho_{\rm DM}/\rho_{\rm c}$, in terms of the DM energy density $\rho_{\rm DM}=n_{\rm DM}\,m_{\rm DM}$ and the critical energy density, $\rho_{\rm c}=3H_0^2/8\pi G\simeq 1.053\times 10^{-5}h^2\text{ GeV}/\text{cm}^3$, where $h$ parametrises the Hubble rate, $H\equiv 100\,h\, {\rm km \,s^{-1} \,Mpc^{-1}}$.  
This should be compared with the experimental measurement provided by the Planck satellite, $\Omega_{\rm DM} h^2=0.1198\pm0.0012$~\cite{Planck:2018vyg}. The relic abundance is reproduced if the DM annihilation cross-section satisfies $\med{\sigma v}=\med{\sigma{v}}_{\rm th} \equiv2.2\times 10^{-26}\text{ cm}^3$/s~\cite{Steigman:2012nb}. 

The total DM annihilation cross-section is obtained summing over all possible annihilation channels,   
\begin{equation}
    \sigma=\sum_{\rm SM}\sigma_{\rm SM}+\sigma_{rr}+\sum_{n=1}\sigma_{rG_n}+\sum_{n=1}\sum_{m\geq n}\sigma_{G_nG_m}\,.
\end{equation}
The first term, $\sigma_{\rm SM}$, corresponds to annihilations of DM into SM particles, ${\rm DM\,DM} \to {\rm SM\,SM}$, which are mediated by the exchange of a radion or a KK graviton in the $s$-channel.
The other terms, $\sigma_{rr}$, $\sigma_{rG_n}$ and $\sigma_{G_nG_m}$, describe the DM annihilations into bulk particles, namely two radions, one radion and one KK graviton, and two KK gravitons. In general, within the Goldberger-Wise stabilisation mechanism \cite{GW1,GW2}, the radion is the lightest bulk particle, $m_r\ll m_{1}$. Thus, it is natural to consider the region of the parameter space where $m_{\rm DM}>m_r$ and DM annihilations into radions are kinematically open. 
Moreover, DM can efficiently annihilate into one radion and one KK graviton (two KK gravitons) if $2m_{\rm DM}>m_{n}+m_r$ ($2m_{\rm DM}>m_{m}+m_{n}$). 

The relevant dynamics occurs around the freeze-out temperature, $T_{\rm FO}\simeq m_{\rm DM}/25$, when the DM particles decouple. It is useful to approximate the centre-of-mass energy of the annihilation processes, $s$, as $s\simeq 4m_{\rm DM}^2$, and compute the cross-section at the leading order in the so-called velocity expansion. However, the cross-section exhibits a series of resonances, corresponding to the exchange of a radion or a KK graviton in the $s$-channel for $\sqrt{s}\simeq 2m_{\rm DM}\simeq m_r\,(m_{n})$. Since the velocity expansion may fail in the neighbourhood of a resonance, we compute the analytic value of $\med{\sigma v}$ using the exact expression from Ref.~\cite{Gondolo:1990dk},\footnote{For a rigorous treatment in the presence of narrow resonances, see Ref.~\cite{Binder:2021bmg}.}
\begin{equation}
    \med{\sigma v}=\frac{1}{8m_{\rm DM}^4TK_2^2(m_{\rm DM}/T)}\int_{4m_{\rm DM}^2}^\infty ds\,(s-4m_{\rm DM}^2)\sqrt{s}\,\sigma(s)\,K_1\left(\frac{\sqrt{s}}{T}\right).
\end{equation}
In the next subsections we discuss in more detail the contribution of the different annihilation channels to the total annihilation cross-section.

\subsection{Annihilations into Standard Model particles}

\begin{figure}[h]
	\centering
\includegraphics[width=0.5\textwidth]{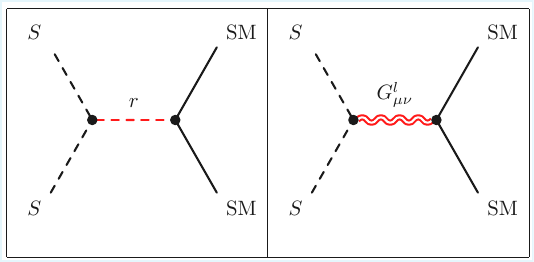}
	\caption{Feynman diagrams for DM annihilations into SM particles.}\label{fig:diagsSM}
\end{figure}

Dark Matter particles unavoidably annihilate into SM particles via the exchange of a radion or a KK graviton in the $s$-channel, see Fig.~\ref{fig:diagsSM}. Depending on the DM mass, different final states can be produced: for $m_{\rm DM}<m_e$, DM can only annihilate into photons, gluons (the latter only before the QCDpt) and neutrinos, while annihilations into a generic $PP$ channel, being $P$ a SM fermion, scalar or gauge boson, are allowed as long as $m_{\rm DM}>m_{\rm P}$.

As discussed in the previous sections, the interactions of SM particles, localized in
the IR brane, with the radion and the KK gravitons are suppressed by the scale $\Lambda_{\rm IR}$. The couplings of DM particles to radions and KK gravitons are suppressed by the energy scale $\Lambda_{\rm DIR}\ll\Lambda_{\rm IR}$. Therefore, the total cross-section into SM states scales as $\sigma_{\rm SM}\propto (\Lambda_{\rm DIR}^2\Lambda_{\rm IR}^2)^{-1}$. We consider values of $\Lambda_{\rm IR}> 10$ TeV to avoid strong constraints from collider searches. We find that DM annihilations into SM particles are typically suppressed compared to those into bulk particles (see discussion in the next subsection). This is particularly evident when $\Lambda_{\rm IR}\gg \Lambda_{\rm DIR}$,
see Fig.~\ref{fig:sigmav}. However, note that for $m_{\rm DM}\simeq \Lambda_{\rm IR}\simeq\Lambda_{\rm DIR}=[5-10]$ TeV the relic abundance is reproduced by DM annihilations into SM particles, in agreement with the standard two-brane RS scenario~\cite{Folgado:2019sgz}. 

\subsection{Annihilations into bulk particles: radions and KK gravitons}

\begin{figure}[h]
	\centering
\includegraphics[width=0.9\textwidth]{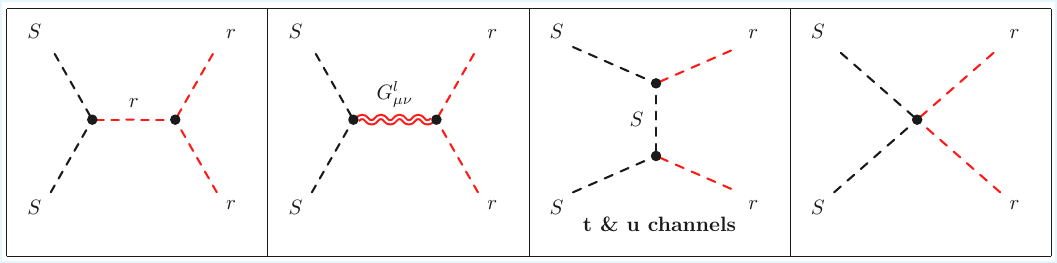}
	\caption{Feynman diagrams for DM annihilations into two radions.}\label{fig:diags2r} 
\end{figure}

\begin{figure}[h]
	\centering
\includegraphics[width=0.9\textwidth]{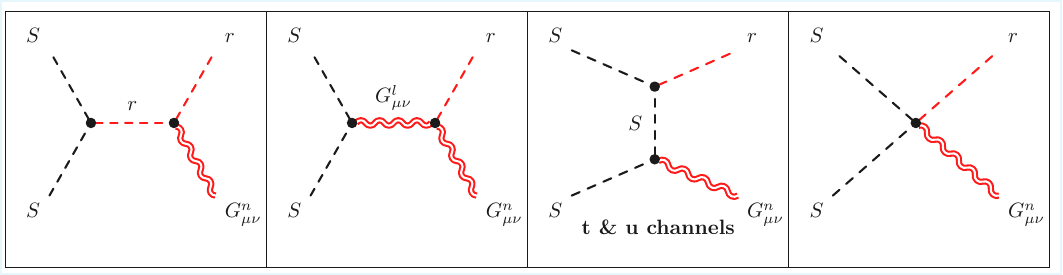}
	\caption{Feynman diagrams for DM annihilations into a radion and a KK graviton.}\label{fig:diagsKKr}
\end{figure}

\begin{figure}[h]
	\centering
\includegraphics[width=0.9\textwidth]{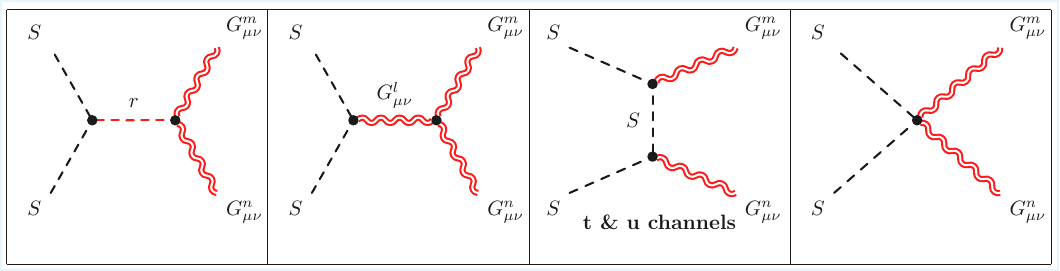}
	\caption{Feynman diagrams for DM annihilations into KK gravitons.}\label{fig:diagsKK}
\end{figure}

Dark Matter particles can also annihilate into bulk particles if the process is kinematically allowed. In the following, we assume $m_{\rm DM}>m_r$, so that ${\rm DM\,DM}\to rr$ is always possible. The Feynman diagrams for this process are shown in Fig.~\ref{fig:diags2r}, while those for annihilations into a radion and a KK graviton, and two KK gravitons, are plotted in Figs.~\ref{fig:diagsKKr} and~\ref{fig:diagsKK}, respectively.

The interactions among DM, radions and gravitons depend only on the scale $\Lambda_{\rm DIR}$. Thus, the cross-section for annihilations into bulk particles scales like $\sigma_{\rm bulk}\propto\Lambda_{\rm DIR}^{-4}$ and it is enhanced with respect to the cross-section into SM particles by a factor $\sigma_{\rm bulk}/\sigma_{\rm SM}\sim (\Lambda_{\rm IR}/\Lambda_{\rm DIR})^2\gg1$. This implies that, whenever it is kinematically allowed, DM mostly annihilates into bulk particles. The cross-section exhibits a series of resonances corresponding to the $s$-channel exchange of a radion or a graviton.

\begin{figure}[htbp] \label{fig:2Dscanall}
    \centering  
\includegraphics[width=1.0\textwidth]{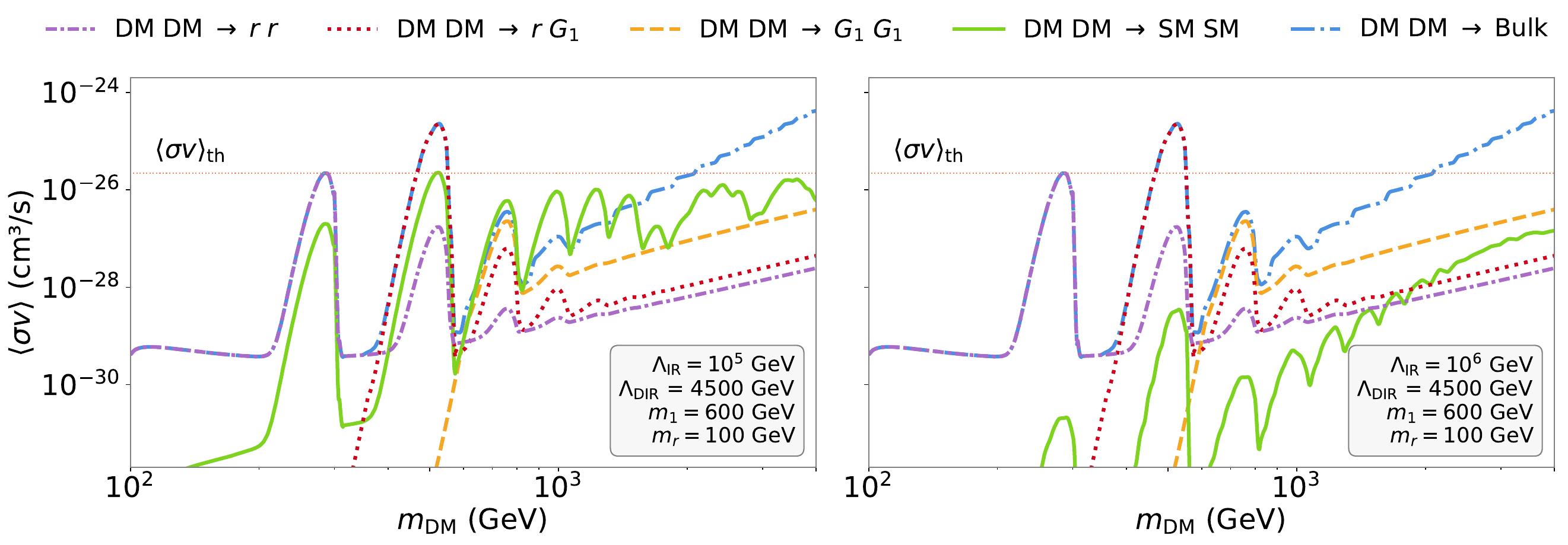}
    \caption{Thermally-averaged annihilation cross-section versus the DM mass for the different production channels: radions, radion plus lightest graviton, lightest gravitons and SM in dotted-dashed purple, dotted red, dashed orange and solid green, respectively, for $\Lambda_{\rm IR}= 10^5$ GeV (\emph{left}) and $\Lambda_{\rm IR}= 10^6$ GeV (\emph{right}). In addition, the sum of all bulk final states is shown in dotted-dashed blue. The freeze-out value of the thermally-averaged cross-section, $\langle \sigma v\rangle_{\rm th}$, is shown as a dotted red horizontal line. }  \label{fig:sigmav}
\end{figure}

In Fig.~\ref{fig:sigmav} we show the thermally-averaged cross-section $\langle \sigma v\rangle$ as a function of the DM mass for $\Lambda_{\rm DIR}=4.5$ TeV, $m_{1}= 600$ GeV, $m_r=100$, and $\Lambda_{\rm IR}=100$ TeV  GeV (left) or $\Lambda_{\rm IR}=1000$ TeV (right). We show different final states: radions, radion plus lightest graviton, lightest gravitons and SM in dotted-dashed purple, dotted red, dashed orange and solid green, respectively. The sum of all bulk final states is shown in dotted-dashed blue. We also highlight the freeze-out value of the thermally-averaged cross-section, $\langle \sigma v\rangle_{\rm th}$, as a dotted red horizontal line. As can be observed, the larger the IR scale, the smaller the relative strength of the annihilation cross-section into SM with respect to that into radions/gravitons. We have verified that, for $\Lambda_{\rm IR}=\Lambda_{\rm DIR}$, annihilations into SM particles dominate the relic abundance. The correct relic abundance is achieved for roughly $m_{\rm DM}=2$ TeV in both plots. The couplings of the KK gravitons depend on the KK number $n$, as discussed in Sec.~\ref{sect:threebranes}. This explains the different heights of the graviton resonances in the SM channel that appear in Fig.~\ref{fig:sigmav}.
 
\section{Theoretical and phenomenological constraints}   \label{sec:bounds} 

\subsection{Consistency of the 4D Effective Field Theory}

 Bulk particles (radion and KK gravitons) become strongly coupled at energies greater than $\Lambda_{\rm DIR}$. In the region of the parameter space in which $2m_{\rm DM}<m_r+m_{1}$ (where we assume $m_r<m_{1}$) the gravitons are never produced as on-shell particles but only enter as virtual particles in the annihilation processes. Thus, we can integrate them out and consider the Effective Field Theory (EFT) without them. In such a case, the consistency of the computation requires that $\Lambda_{\rm DIR}>\max[m_{\rm DM},m_r]$. On the other hand, if the $n-$th KK graviton can be produced on-shell, one must include it as a low-energy degree of freedom, so that the consistency condition becomes stronger, $\Lambda_{\rm DIR}>\max[m_{n},m_{\rm DM}]$. To summarize, we require that $\Lambda_{\rm DIR}>m_{\rm max}$, being $m_{\rm max}$ the heaviest particle that can be produced on-shell (including the DM).
Regarding the unitarity condition of the EFT, which we estimate as $\sigma < 1/s \simeq 1/(2 m_{\rm DM})^2$, we have checked that it is satisfied in all the range of DM masses considered, i.e., [10 GeV, 10 TeV].

\subsection{Direct detection limits}
Direct detection (DD) experiments search for elastic scatterings of DM particles off target nuclei. They provide strong constraints for DM masses in the GeV-TeV range. The zero-momentum DM-nucleus cross-section is parametrised as  
\begin{equation}
		\sigma^A_{\rm SI}=\frac{\mu^2_A}{\mu^2_p}\,(Z+(A-Z)f_n/f_p)^2\,\sigma^p_{\rm SI}, \qquad \text{with}\qquad \sigma^p_{\rm SI}\equiv \frac{\mu^2_p}{\pi}\,f_p^2,
	\end{equation}
where $Z(A)$ are the number of protons and the mass number of the target nuclei, $f_{p,n}$ are the DM couplings to protons and neutrons, $\mu_A\equiv m_A m_{\rm DM}/(m_A+m_{\rm DM})$ is the DM-nucleus reduced mass and $\mu_p\equiv m_p m_{\rm DM}/(m_p+m_{\rm DM})$ is the DM-proton reduced mass.

At the fundamental level, the elastic scattering of DM off a nucleus is mediated by the exchange of either a radion or a graviton in the $t$-channel. 
The radion interacts with both the SM and the DM particles through the trace of the energy-momentum tensor, $\mathcal{L}_{\rm int}\propto r\, T_{{\rm DM\,(SM)}}/\Lambda_{\rm DIR\,(IR)}$, see Eq.~\ref{eq:radEFT}. 
These interactions can be matched onto the low-energy EFT for DM-nucleus scattering, in particular to the scalar operators
\begin{equation}
	\mathcal{O}_q=c_q^rm_q S^2\bar{q}q,\,\,\qquad \mathcal{O}_g=c_g^r\frac{\alpha_s}{\pi}G_{\mu\nu}^aG_a^{\mu\nu}S^2.
\end{equation}
The first operator parametrises DM-quark interactions, while the second one corresponds to DM-gluon interactions, arising at one loop. The Wilson coefficients $c_q^r$ and $c_g^r$ are obtained by matching the amplitudes. We find (see App.~\ref{app:DD} for more details)
\begin{equation}
	c_q^r=\frac{m^2_{\rm DM}}{6m_r^2\Lambda_{\rm IR}\Lambda_{\rm DIR}}\,,\,\, \qquad c_g^r=\frac{C_3\,m^2_{\rm DM}}{48m_r^2\Lambda_{\rm IR}\Lambda_{\rm DIR}}\,.
\end{equation}
As expected, DD experiments are sensitive only to the combination of scales $\sqrt{\Lambda_{\rm DIR}\,\Lambda_{\rm IR}}$.
The relevant nucleon matrix elements are given by 
\begin{equation}
\braket{N|m_q\Bar{q}q|N}=m_Nf_{T_q}^N, \qquad  \braket{N|\frac{\alpha_s}{\pi}G_{\mu\nu}^aG^{\mu\nu}_a|N}\simeq -\frac{8}{9}m_N(1-\sum_{q=u,d,s}f_{T_q}^N)\,,
\end{equation}
where $f_{T_q}^N$ are the mass fractions of light quarks $q$ in nucleon $N=n,p$: $f_{T_u}^p=0.023$, $f_{T_d}^p=0.032$ and $f_{T_s}^p=0.020$ for the proton and $f_{T_u}^n=0.017$ $f_{T_d}^n=0.041$ and $f_{T_s}^n=0.020$ for the neutron \cite{Hisano:2015bma,DelNobile:2021wmp}.
Finally, the nucleon couplings are \cite{Hisano:2015bma,DelNobile:2021wmp}
\begin{equation}\label{eq:DDradion}
	 f_N^r/m_N=\frac{1}{m_{\rm DM}}\left(\sum_{q=u,d,s}c_q^rf_{T_q}^N-\frac{8}{9}\hat{C}_g^r(1-\sum_{q=u,d,s}f_{T_q}^N)\right),
\end{equation}
with 
\begin{equation}
	\hat{C}_g^r=c_g^r-\frac{1}{12}\sum_{q=c,b,t}c_q^r\,,
\end{equation}
where the second term takes into account the contribution of heavy quarks which interact at one loop with the gluons inside the nucleons.

For DM masses in the GeV-TeV range, the strongest constraints on the DM-nucleus cross section come from the LUX-ZEPLIN (LZ) collaboration~\cite{LZ:2024zvo}.  Future experiments, such as DARWIN~\cite{DARWIN:2020jme} and DarkSide-20k~\cite{DarkSide-20k:2017zyg}, are expected to improve the current bounds (see also Ref.~\cite{Billard:2021uyg} for a comprehensive review).
Given that $\sigma_{\rm SI}^p\propto 1/m_r^4$, for any given value of $m_{\rm DM}$, the experimental bounds on the cross-section set a lower bound on the radion mass.

Graviton interactions with matter are proportional to the energy-momentum tensor, $h_{\mu\nu}^n(T^{\mu\nu}_{\rm IR}/\Lambda_{\rm IR}^n +T^{\mu\nu}_{\rm DIR}/\Lambda_{\rm DIR})$, see Eq.~\ref{eq:gEFT}. The latter can be decomposed as the sum of its trace, $T$, and its trace-less part, $\widetilde{T}_{\mu\nu}$. The interaction of a graviton with the trace $T$ is analogous to the interaction of the radion and thus can be mapped at low energies to the same scalar operators, after computing the appropriate Wilson coefficients. On the other hand, the interaction with the trace-less part of the energy-momentum tensor can be matched at low energies to the so-called 
\textit{spin-2 twist-2 operators} for quarks and gluons~\cite{Hisano:2015bma},
\begin{equation}
		O_{\mu\nu}^q=\frac{i}{2}\bar{q}\left(D_\mu\gamma_\nu+D_\nu\gamma_\mu-\frac{1}{2}\eta_{\mu\nu}\slashed{D}\right)q\,,\qquad
		O_{\mu\nu}^g=G_{\mu}^{A\rho}G_{\nu \rho}^A-\frac{1}{4}\eta_{\mu\nu}G^A_{\rho\sigma}G_A^{\rho\sigma}\,.
\end{equation}
The nucleon matrix elements for these operators depend on the second moments of the Parton Distribution Functions (PDFs), 
\begin{equation}
	q(2,\mu)=\int_{0}^{1}dx x f_q(x,\mu)\,,
\end{equation}
for quarks, and analogously for antiquarks $\bar{q}(2,\mu)$ and gluons $G(2,\mu)$.
The DM couplings to nucleons can thus be computed as~\cite{Carrillo-Monteverde:2018phy, Lee:2024wes}
\begin{equation}
	\begin{split}
	f_N^{n}/m_N=\frac{m_{\rm DM}}{4m_{n}^2\Lambda_{\rm IR}^n\Lambda_{\rm DIR}}&\bigg[\sum_{q=u,d,s,c,b}3[{q}(2,m_{n})+\bar{q}(2,m_{n})]+3\,G(2,m_{n})\\
	&+\sum_{q=u,d,s}\frac{1}{3}f_{T_q}^N-\left(\frac{2}{27}-\frac{C_3}{27}\right)(1-\sum_{q=u,d,s}f_{T_q}^N)
	\,\bigg],	
    \end{split}
\end{equation}
We consider the contribution of both radion and gravitons to the cross-section in our numerical computations. The contribution to the  cross-section from radion exchange is inversely proportional to $(m_r^2\,\Lambda_{\rm IR})^2$, while the one from graviton exchange is inversely proportional to $(m_{n}^2\Lambda_{\rm IR}^n)^2$. As long as the gravitons are heavy enough, so that $m_r^2\,\Lambda_{\rm IR}\ll m_{n}^2\Lambda_{\rm IR}^n$, the cross-section is  dominated by the exchange of a radion. This is always satisfied in the left plots of Fig.~\ref{fig:2Dscanr}, as the gravitons are very heavy. On the other hand, lighter gravitons can contribute at the same level or even more than the radion. This can potentially affect part of the parameter space of right plots of Fig.~\ref{fig:2Dscanr}, in which, however, DD constraints are not strong enough to exclude any region. More details are provided in App.~\ref{app:DD}.

\subsection{Indirect detection and Cosmic Microwave Background constraints}\label{sec:indirect}  
Indirect detection (ID) experiments aim to detect spectra of SM particles emitted by DM decays and annihilations. The most promising targets for detecting such signals are the Galactic center and dwarf spheroidal galaxies, which are small satellites of the Milky Way whose dynamics is dominated by their DM component. These searches depend of the combination $n_{\rm DM}^2\med{\sigma v}$, being $n_{\rm DM}$ the DM number density in the target object, and allow to put stringent constraints on the DM annihilation cross-section. Concurrently, competitive constraints (especially for sub-GeV DM masses) can be derived from the analysis of the anisotropies of the Cosmic Microwave Background (CMB). If two DM particles annihilate around $T\sim 0.1$ eV, the injected particles can reionize hydrogen, modifying the evolution of recombination and leaving an imprint on the CMB. The relevant quantity in this case is the total injected power which, for an annihilation process, depends on $\rho_{\rm DM}^2\med{\sigma v}/m_{\rm DM}$. A review of current constraints on the DM annihilation cross-section as a function of the DM mass from ID searches and CMB anisotropies, which we denote by $\med{\sigma v}_{\rm limit}^{m_{\rm DM}}$, can be found in Refs.~\cite{Cirelli:2024kph,Cirelli:2024ssz}. More precisely, limits are derived on the quantity $n_{\rm DM}^2(m_{\rm DM})\med{\sigma v}_{\rm limit}^{m_{\rm DM}}\propto \med{\sigma v}_{\rm limit}^{m_{\rm DM}}/m_{\rm DM}^2$, 
if we assume that the total DM energy density is reproduced by a particle of mass $m_{\rm DM}$.

Concerning the direct annihilation of DM into SM particles, DM DM$\to$ SM SM, we can extract the experimental limit for each channel ${\rm SM\, 
 SM}=\{\bar{b}b,\,\pi\pi,\, W^+W^-,\,\dots\}$  in a straightforward way. However, in the parameter space of our interest, direct annihilations into SM particles are suppressed (with cross-sections much smaller than the current upper limits), and DM dominantly annihilates into bulk particles, so that the analysis is slightly more involved. 
 
For the sake of clarity, we first focus on the region of the parameter space in which DM mainly annihilates into radions via DM DM$\to rr$, so that $\med{\sigma_{rr}v}\simeq\med{\sigma v}_{\rm th}$. The radions produced by the annihilation processes carry an energy $E_r\simeq m_{\rm DM}$. Subsequently, they undergo fast decays into pairs of SM particles, each with energy $E_{\rm SM}\simeq E_r/2\simeq m_{\rm DM}/2$. 
In summary, a flux of SM particles $\Phi(E_{\rm SM})$ is produced from DM annihilations through an intermediate step of radions, which reduces the energy $E_{\rm SM}$ compared to the case of direct annihilations while increasing the number of SM particles produced. 

Although a detailed analysis of the energy spectra is beyond the scope of our work, we perform a simplified analysis to  
estimate the relevant constraints. First, for each SM final state, we re-scale the DM annihilation cross-section by multiplying it by the corresponding BR, $\med{\sigma_{\rm SM}v}_{\rm eff}\equiv\med{\sigma_{rr} v}{\rm Br}(r\to {\rm SM})$. Then, we compare the re-scaled cross-section with the current experimental limit corresponding to that specific SM annihilation channel.  In this second step, we take into account that the flux of SM particles produced by the decay of the radions has roughly an energy $E_{\rm SM}\simeq m_{\rm DM}/2$. The flux itself is proportional to $n_{\rm DM}^2\med{\sigma_{\rm SM} v}_{\rm eff}\propto \med{\sigma_{\rm SM} v}_{\rm eff}/m_{\rm DM}^2$, under the assumption that the observed DM relic abundance is reproduced by a single particle of mass $m_{\rm DM} $. 
The experimental limit on the cross-section obtained for a flux of energy $m_{\rm DM}/2$ is given by $\med{\sigma v}_{\rm limit}^{m_{\rm DM}/2}$. The subtle point is that this value is obtained assuming the number density of a DM particle of mass $m_{\rm DM}/2$, i.e., the actual bound on the flux applies on the quantity $\med{\sigma v}_{\rm limit}^{m_{\rm DM}/2}/(m_{\rm DM}/2)^2$, while our candidate has mass $m_{\rm DM}$. To take into account this fact, we use that the flux of the DM particle of mass $m_{\rm DM}$ scales as $\Phi(E_{\rm SM})\propto \med{\sigma_{\rm SM}v}_{\rm eff}/m_{\rm DM}^2 <\med{\sigma v}_{\rm limit}^{m_{\rm DM}/2}/(m_{\rm DM}/2)^2$. This implies the constraint $\med{\sigma_{\rm SM}v}_{\rm eff}<4\,\med{\sigma v}_{\rm limit}^{m_{\rm DM}/2}$ for a DM particle of mass $m_{\rm DM}$. Thus, for a given DM mass, the limit is 4 times weaker compared to the one extracted by direct annihilation into SM particles. Analogously, limits stemming from CMB anisotropies are obtained as $\med{\sigma_{\rm SM}v}_{\rm eff}<2\,\med{\sigma v}_{\rm limit}^{m_{\rm DM}/2}$. 

Radions heavier than a few GeV mostly decay into heavy quarks (mainly $\bar{c}c$ or $\bar{b}b$), light mesons and EW gauge bosons (if $m_r\gtrsim 2m_W$), which subsequently produce an electromagnetic cascade. This is constrained by gamma ray data from the Fermi-LAT collaboration~\cite{Fermi-LAT:2015att}, which excludes DM candidates with annihilation cross-sections that produce SM fluxes with energies up to $E_{\rm SM}\sim 200$ GeV. Notice that if the radion dominantly decays into light mesons, the constraints are significantly weaker. In particular, ID searches do not exclude regions of the parameter space with $m_r\lesssim 2m_c$. Furthermore, for $2m_\tau\lesssim m_r\lesssim 2m_b$ and  $50\text{ GeV}\lesssim m_r\lesssim 2m_W$  the BR to heavy quarks decreases thus providing weaker constraints. 
This explains the little hole in the orange region of Fig.~\ref{fig:2Dscanr}, as well as the fact that the bound gets weaker for $m_r\gtrsim 50$ GeV. The Cherenkov Telescope Array (CTA) is expected to improve the current bounds  for DM masses in the [1-10] TeV range~\cite{Morselli:2017ree}, which in our case is already excluded by DD null results.

More generally, DM also annihilates into KK gravitons. Given that $m_{1}\gg m_r$ and $\Lambda_{\rm IR}\gg \Lambda_{\rm DIR}$, heavier gravitons dominantly decay into lighter gravitons,
$G_m\to G_n\, G_l$ ($n>m,l\geq1$), while the lightest graviton, $G_1$, mostly decays into radions, $G_1\to r\,r$, which eventually disintegrates into SM particles. This provides a complicated chain of decays, where multiple intermediate steps produce a flux of SM particles, 
generalizing the picture described before. Once again, a detailed calculation of the energy spectra for these multi-step decays is beyond the scope of this work. Generalizing the previous discussion, the annihilation of DM into SM particles generically requires $n$ intermediate steps, each one reducing the energy of the SM final products. At the final stage of the annihilations we are left with $\sim 2^n$ SM particles, each one carrying roughly an energy $E_{\rm SM}\sim (m_{\rm DM}/2^n)$. The constraints become weaker as the number of intermediate steps, $n$, increases.

 \subsection{LHC bounds}

\begin{figure}[htbp]
	\centering
\includegraphics[width=1\textwidth]{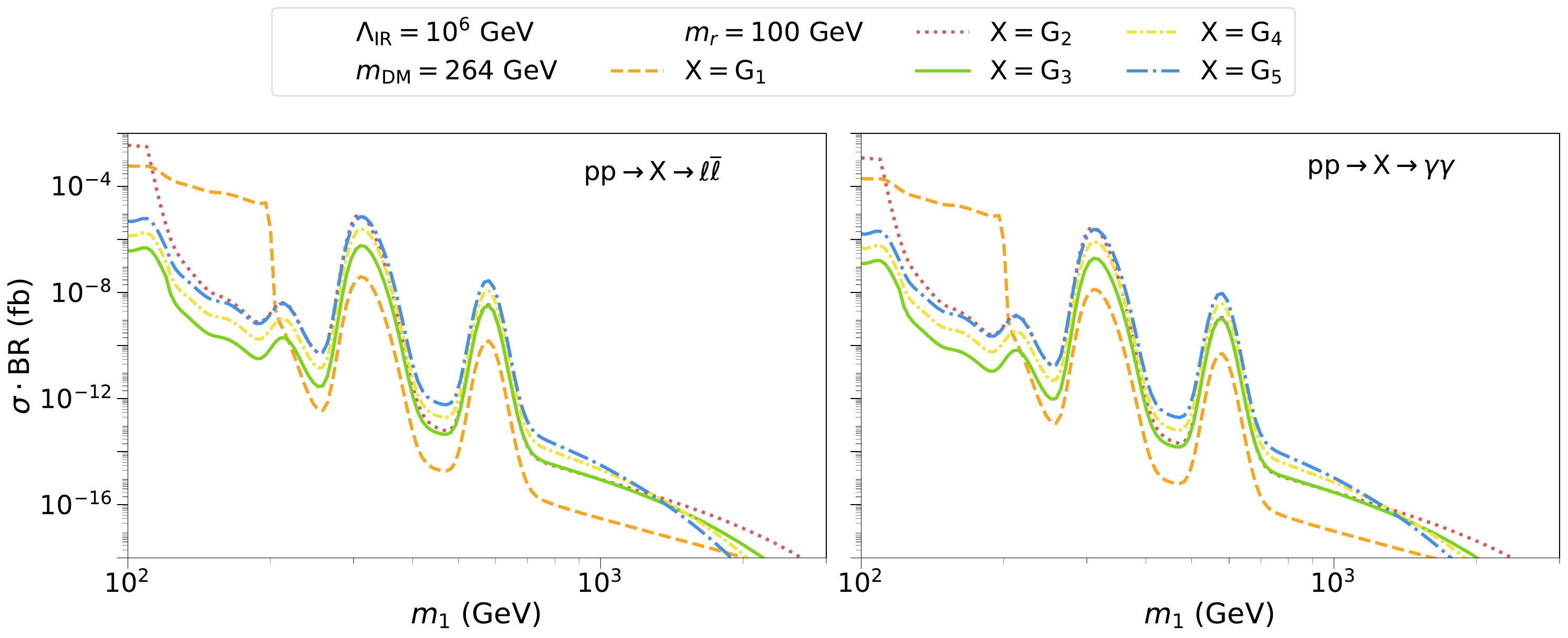}
	\caption{Production cross section of di-lepton \textit{(left)} and di-photon  \textit{(right)} channels for the first five KK gravitons.  For each value of $m_1$, we select the value of $\Lambda_\text{DIR}$  such that the DM relic abundance is reproduced. The three peaks correspond to the $n= 3, 2, 1$ KK resonances, respectively. As can be observed in Fig.~\ref{fig:2Dscanr}, in order to reproduce the relic abundance at a resonance, $\Lambda_\text{DIR}$ must increase considerably. This results in an increase of the BRs into SM particles because $\Lambda_\text{DIR}$ and $\Lambda_\text{IR}^{n}$ are closer to each other. The masses of the KK gravitons are given by $m_2 \simeq  1.83$ $m_1$, $m_3 \simeq 2.66$ $m_1$, $m_4 \simeq 3.48$ $m_1$ and $m_5 \simeq 4.3$ $m_1$. Although both figures are very similar, it can be seen that the values in the figure on the left are larger than those on the right, since the BRs into leptons are always larger than into photons (see Fig.~\ref{fig:decaysG}.)}\label{fig:LHCcross}
\end{figure}

Resonance searches at the LHC may provide constraints on the parameter space of the model. Indeed, both the radion and the first KK graviton modes could be resonantly produced through quark and gluon fusion at hadron colliders. The production cross-section of the $n$-KK graviton at the LHC is given by~\cite{Folgado:2019sgz}
\begin{equation}\label{eq:Gprod}
 \sigma_{pp\to G_n}(m_{n})=\frac{\pi}{48(\Lambda_{\rm IR}^n)^2}\left[3\mathcal{L}_{gg}(m_{n}^2)+\sum_q 4\mathcal{L}_{q\bar{q}}(m_{n}^2)\right],
\end{equation}
where
\begin{equation}
	\mathcal{L}_{ij}(\hat{s})=\frac{\hat{s}}{s}\int_{\hat{s}/s}^1\frac{dx}{x}f_i(x)f_j\left(\frac{\hat{s}}{xs}\right)
\end{equation}
are the luminosity functions, with $f_i$ being the PDFs evaluated at $Q^2=m_{n}^2$. The production cross-section of a radion at the LHC is instead dominated by gluon-gluon fusion,
\begin{equation}\label{eq:rprod}
    \sigma_{pp\to r}(m_r)=\frac{\alpha_s^2C_3^2}{1536\pi\Lambda_{\rm IR}^2}\mathcal{L}_{gg}(m_{r}^2)\,.
\end{equation}
The expressions in Eqs.~\eqref{eq:Gprod} and~\eqref{eq:rprod} are analogous to those of Ref.~\cite{Folgado:2019sgz} for the standard two-brane RS scenario, with a crucial difference: the production cross-section of the $n-$th KK graviton is suppressed by $(\Lambda_{\rm IR}^n)^2$, while the one of the radion is suppressed by $\Lambda_{\rm IR}^2$ (in a two-brane setup, both cross-sections are suppressed by the common scale $\Lambda_{\rm IR,2b}^2$~\cite{Folgado:2019sgz}). This reflects the nature of our three-brane setup, in which the KK gravitons and the radion interact with SM matter with different couplings. Since $\Lambda_{\rm IR}\ll\Lambda_{\rm IR}^{n=1}$, the resonant production of radions can be dominant over the production of $G_1$, particularly when $\Lambda_{\rm IR}$ is orders of magnitude larger than $\Lambda_{\rm DIR}$ (contrary to the two-brane scenario, see Ref.~\cite{Folgado:2019sgz}). On the other hand, higher KK graviton modes can have stronger couplings to the SM, $\Lambda_{\rm IR}^n<\Lambda_{\rm IR}$, especially when $\Lambda_{\rm DIR}\lesssim \Lambda_{\rm IR}$, see Fig.~\ref{fig:KKcouplingsa}. This can enhance their production cross-section, which is, however, simultaneously suppressed by their large mass. The production cross-section (times BR to leptons or photons) for the first 5 KK gravitons is shown in Fig.~\ref{fig:LHCcross} as a function of the mass of the first mode, $m_{1}$. We can observe that, for some value of $m_{1}$, higher KK modes can be produced with a larger rate.

The prediction for the shape of the graviton resonances is therefore qualitatively different from the usual two-brane scenario, where all the gravitons couple with the same strength to SM particles.
This
opens an intriguing possibility: if a collider experiment is ever able to observe multiple
resonances, one could use the location, width, and peak value of the first two observed
resonances (either the radion and the first graviton or the first 2 gravitons) to fix the value
of the parameters of the model, namely $\{m_{1} , m_r, \Lambda_{\rm DIR}, \Lambda_{\rm IR}\}$. Then, the properties of the
following resonances are a prediction of the model, which would also allow to distinguish our
three-brane scenario with the SM in the intermediate brane from the usual two-brane scenario, thus providing a smoking-gun signature
of models with multiple branes\footnote{Notice that the observation of KK graviton resonances with mode-dependent couplings to SM particles would imply that the SM cannot lie on the DIR brane, as in such a case the couplings would be universal. Furthermore, since the main assumption of the Randall-Sundrum framework is that the fundamental scale is the Planck scale, a particle living on the UV brane ($y=0$) interacts with bulk particles (radion and KK gravitons) with Planck-suppressed couplings.
Thus, if the SM lived on that brane, the LHC would not be able to observe KK graviton resonances. Therefore, a positive signal would imply that the SM is not localized on the DIR brane nor on the UV one but on a another intermediate one.}.

Once the graviton (or the radion) is produced on-shell inside the collider, it immediately decays into SM particles, so that one can extract experimental limits on $\sigma(pp\to{\rm SM})=\sigma_{pp\to G_n}\times {\rm BR}(G_n\to {\rm SM})$, and analogously for the radion. The more stringent constraints are provided by di-photon ($pp\to \gamma\gamma$)~\cite{ATLAS:2021uiz} and di-lepton ($pp\to l\bar{l}$)~\cite{ATLAS:2019erb} searches. These limits apply to both spin-0 and spin-2 resonances with masses $\gtrsim 300$ GeV. This covers most of our parameter space in graviton mass, $m_{1}$, while we typically consider lighter radion masses. The radion interacts with SM matter in the same way as the Higgs boson with coupling re-scaled by $\theta=v_{\rm EW}/\sqrt{6}\Lambda_{\rm IR}\simeq (100\text{ GeV}/\Lambda_{\rm IR})$. Ref.~\cite{ATLAS:2024bjr} discusses the limits for a scalar particle which couples to matter proportionally to the mass like the Higgs boson, such as our radion.
Ref.~\cite{Csaki:2000zn} discusses the constraints on radion masses in the $10-100$ GeV range from LEP, concluding that no limits can be obtained for $\Lambda_{\rm IR}\gtrsim 1$ TeV. Radions with masses $300\,\text{MeV}\lesssim m_r\lesssim 5$ GeV are constrained by displaced vertex searches of $B$-meson decays $B\to Kr(\mu\mu)$ at LHCb, see Refs.~\cite{LHCb:2015nkv,LHCb:2016awg} and Ref.~\cite{LHCb:2025rin} for future sensitivities (the limits are expressed in term of the parameter $\theta$ defined above, usually interpreted as a mixing angle between the SM Higgs boson with the new scalar particle).    Current limits exclude radion masses $300\text{ MeV}\lesssim m_r\lesssim 5$ GeV for $\Lambda_{\rm IR}\lesssim 300$ TeV. Notice that this region of the parameter space is already excluded by DD and ID constraints. Lighter radions are more severely constrained by beam-dump experiments and stellar limits, and we do not consider this mass range.

An additional process which could take place at colliders is a monojet + missing energy event: two protons producing a pair of DM particles (missing energy) and one jet, through the $s-$channel exchange of a radion or a graviton. As discussed in Ref.~\cite{Bai:2010hh}, these events allow us to put limits on the DM-nucleus elastic cross-section, $\sigma_{\rm SI}$. However, for the range of DM masses that we consider in this work, $m_{\rm DM}\gtrsim 10$ GeV, the limits from colliders are not competitive with those provided by DD experiments.\footnote{To the best of our knowledge, no  computations for monojet emission from twist-2 operators exist in the literature.}

\section{Numerical analysis} \label{sec:results}
The relevant independent parameters of the model are the IR scale $\Lambda_{\rm IR}$, the DIR scale $\Lambda_{\rm DIR}$, the DM mass $m_{\rm DM}$, the first KK graviton mass $m_{1}$ and the radion mass $m_{r}$. We work in the limit of small brane backreaction, such that one of the radions is light and the other one is very heavy. Furthermore, in this case the natural mass hierarchy is $m_{1}>m_{r}$, which is the regime we consider in this work. In order to study the allowed parameter space of the model that correctly reproduces the DM relic abundance and is compatible with the theoretical and experimental constraints, we perform a numerical scan. The FeynRules package \cite{Alloul_2014,Christensen_2009} has been used to obtain the vertices and the FeynCalc package \cite{Shtabovenko_2025,Shtabovenko_2020, Shtabovenko_2016, Mertig_1991} has been used to calculate the different cross-sections. Subsequently, each of the latter have been integrated numerically to obtain their thermally-averaged values, $\langle \sigma v\rangle$. The $\Lambda_{\rm DIR}$ that reproduces the DM relic abundance is obtained by varying it in the range $[m_{\rm DM}, \Lambda_{\rm IR}]$ and solving Eq.~\ref{eq.BEQ}. This procedure gives the
abundance with an accuracy of less than 1\%. We fix the values of $\Lambda_{\rm IR}$ and $m_{1}$, and obtain the value of $\Lambda_{\rm DIR}$ that reproduces the relic abundance as a function of $m_{\rm DM}$ and $m_{r}$. We impose constraints from DD, ID and the LHC, as well as require that the EFT approach is valid, i.e., $\Lambda_{\rm DIR}>\max[m_{\rm DM},m_{r},m_n]$. 

Three different cases are possible, depending on the mass hierarchy of the DM particles, the radions, and the KK gravitons: 
\begin{itemize}
\item[\emph{i)}] $m_{1}+m_r> 2 m_{\rm DM}> 2 m_r$: only annihilations into radions are kinematically allowed.
\item[\emph{ii)}] $ 2 m_{1}> 2 m_{\rm DM}> m_{n}+m_r$: annihilation channels into radions and a radion plus KK gravitons are open.
\item[\emph{iii)}] $  m_{\rm DM}>  m_{1}$: all annihilation channels are open.
\end{itemize}

\subsection{Annihilations into radions and radions plus KK gravitons}

\begin{figure}[htbp]
	\centering
\includegraphics[width=1\textwidth]{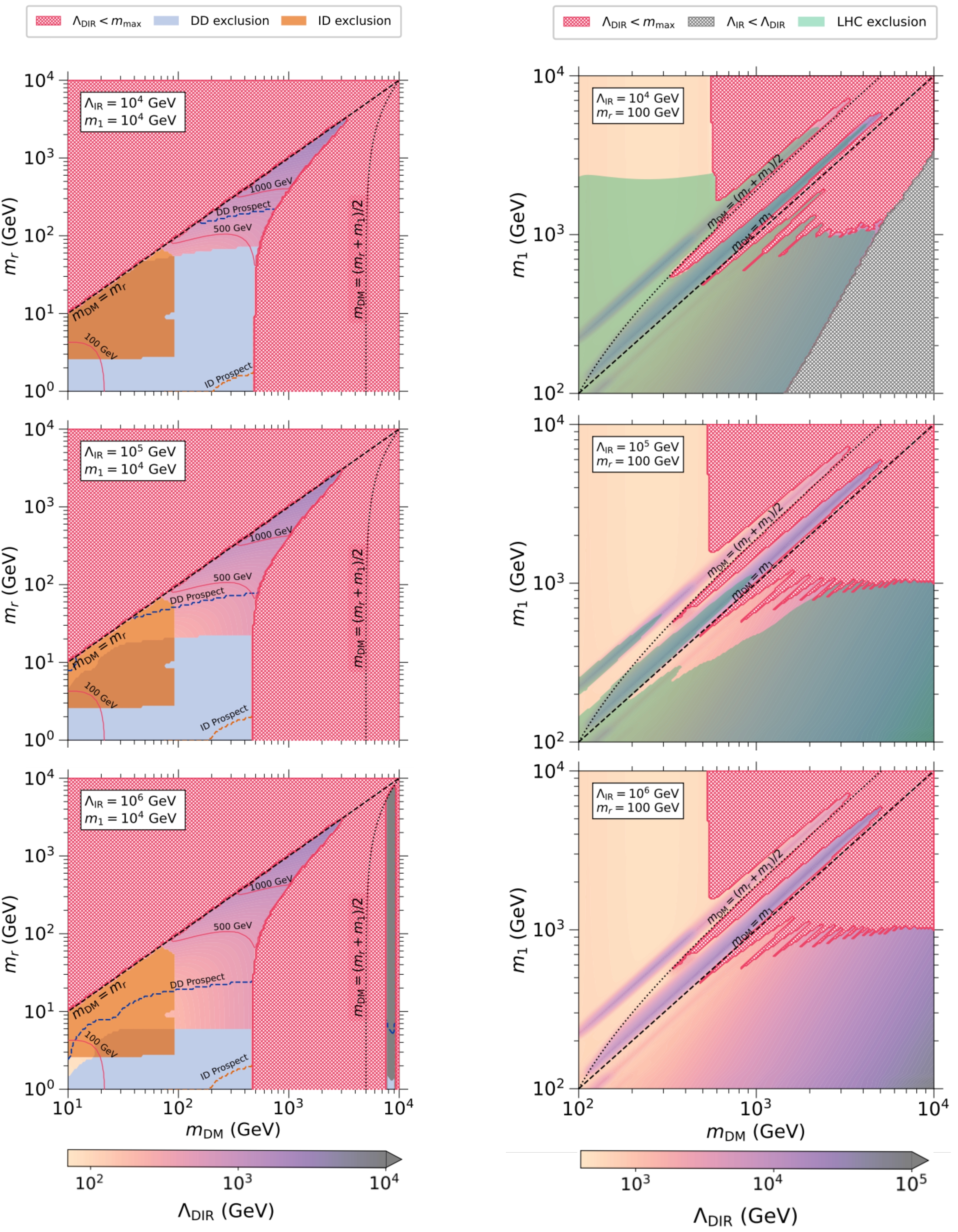}
	\caption{
    Results on the plane of the DM mass and the radion (first KK graviton) mass in the \emph{left} (\emph{right}) plots for fixed KK graviton masses (radion mass) and $\Lambda_{\rm IR}$. In every point the relic abundance is reproduced for a different value of $\Lambda_{\rm DIR}$, shown with the color map and some contour lines. Channels into two KK gravitons are kinematically open only for the plots on the right. The blue and orange dashed lines correspond to the prospects for DD and ID, respectively.
}\label{fig:2Dscanr}
\end{figure}

In this section we focus on the parameter space where the DM annihilation channel into two KK gravitons is closed, for which the most relevant parameters are $m_r, m_{\rm DM}$ and $\Lambda_{\rm DIR}$.
In the left plots of Fig.~\ref{fig:2Dscanr} we depict the radion mass versus the DM mass for $m_{1}=10$ TeV and $\Lambda_{\rm IR}=10,\,100,\,1000$ TeV.  In every point, the relic abundance is reproduced for a different value of $\Lambda_{\rm DIR}$, as the color code (depicted below the bottom panels) indicates. We also highlight some  contours of $\Lambda_{\rm DIR}$ as a guideline.
The red region is excluded by consistency of the EFT. The blue and orange shaded regions are excluded by DD and ID constraints, respectively. As one observes, only a small region of the parameter space survives all constraints, with roughly $m_{\rm DM}\simeq [0.1-4]$ TeV, and  $m_{r}\simeq [0.005-30]$ TeV, depending on the value of $\Lambda_{\rm IR}$. Notice that, whenever the $G_n\,r$ channel is open, it dominates over the $r\,r$ channel.

Note that the bounds provided by ID do not change when we vary $\Lambda_{\rm IR}$ or $\Lambda_{\rm DIR}$, because they only depend on the DM mass and the radion decay BRs into SM particles; the latter are independent of both scales, since the radion interacts with
all species living in the IR brane  with universal coupling $1/{\Lambda_{\rm IR}}$. 
On the other hand, the DD bound depends on both $\Lambda^{ n}_{\rm IR}$ and $\Lambda_{\rm DIR}$. Note that by increasing $\Lambda_{\rm IR}$, and therefore $\Lambda^{ n}_{\rm IR}$, the DD bound weakens significantly, specially for $\Lambda_{\rm IR} \gg \Lambda_{\rm DIR}$. It is also interesting to point out that the $\Lambda_{\rm DIR}$ values for which the relic abundance
is achieved are practically the same for the three cases of $\Lambda_{\rm IR}$ considered. This is because the relic abundance is mostly produced by the radion channel, which only depends on $\Lambda_{\rm DIR}$, with the SM particles channel not playing any relevant role.

In the particular case of $\Lambda_{\rm IR} = 10^{6}$ GeV
(bottom left plot) we can observe the resonance of the second KK graviton (vertical gray strip) once the $G_1\,r$ channel opens up. Due to the considerable growth of the cross section at the peak of the resonance, in order to reproduce the DM relic abundance, a compensating growth of $\Lambda_{\rm DIR}$ is needed. It is for this reason that there is a large difference between the $\Lambda_{\rm DIR}$ values in this region and those in the rest of the plot.

 \subsection{Annihilations into radions and KK gravitons}

Now we consider the case in which all bulk states (radion and KK gravitons) are kinematically accessible. In this regime, it is interesting to study the dependence on $m_{1}$; therefore, in the right plots of Fig.~\ref{fig:2Dscanr}  we show the plane of the first KK graviton mass versus the DM mass for fixed values $m_r=100$ GeV and $\Lambda_{\rm IR}=10,\,100,\,1000$ TeV. The red hashed region is excluded by the validity of the EFT, while the green region is excluded by LHC bounds. The gray hashed regions is excluded because $\Lambda_{\rm IR} < \Lambda_{\rm DIR}$. The colored density regions show the value of $0.5\,{\rm TeV}<\Lambda_{\rm DIR}<100\,{\rm TeV}$ for which the relic abundance is reproduced, with larger values corresponding to larger values of $m_{\rm DM}$. Contours where annihilations into a radion and a graviton, and into gravitons, open up are shown with dotted and dashed black lines, respectively. Kaluza-Klein resonances are clearly visible to the right of the dashed black line.

It can be seen that the value of $\Lambda_{\rm DIR}$ remains roughly unchanged when the channel into radions is the only one open, which is due to the fact that the cross-section to KK gravitons does not come into play and therefore their mass cannot have any effect, except on the resonance of the first one. We have chosen a radion mass which satisfies DD and ID constraints for the three values of $\Lambda_{\rm IR}$ considered. When the $G_1\,r$ channel opens up, we find a second resonance (note that at the resonances, $\Lambda_{\rm DIR}$ must increase to reproduce the relic abundance). Similarly, once the channels into KK gravitons open up, $\Lambda_{\rm DIR}$ must increase accordingly.

LHC bounds become more restrictive when the value of $\Lambda_{\rm IR}$ decreases, because the production cross-section of radions (KK gravitons) depends on the inverse of the square of $\Lambda_{\rm IR}$ ($\Lambda_{\rm IR}^{n}$). On the other hand, the effect of $\Lambda_{\rm DIR}$ is the opposite. The closer $\Lambda_{\rm DIR}$ gets to $\Lambda_{\rm IR}$, the larger the BRs of the bulk particles into the SM, which result in a more restrictive bound.
 
Finally, in Fig.~\ref{fig:2Dscanall} we present a plot showing the parameter space of $\Lambda_{\rm DIR}$ and $m_\text{DM}$, for fixed $m_{1}=420$ GeV and $m_r=100$ GeV. We show $\Lambda_{\rm IR}=100 \,(1000)$ TeV in the left (right) panel. As in the previous plots, the EFT is not valid in the hatched red region,\footnote{The different regimes may be understood as follows: on the one hand, when the DM is light enough such that it can only produce SM particles or radions, the heaviest particle that plays a role is the DM itself, so its mass is the limiting one, $m_{\rm max}\simeq m_{\rm DM}$. On the other hand, when $m_{\rm DM}\gg m_1$, it is able to produce very massive KK gravitons due to the opening of multiple $G_nr$ channels, with $m_n>m_{\rm DM}$, so that the heaviest particle which sets the limit is $m_{\rm max}\simeq 2m_{\rm DM}-m_r \simeq 2m_{\rm DM}$.
} and the LHC limits exclude the green area. The correct DM relic abundance is obtained along the solid black line. The relic abundance is provided dominantly by annihilations into radions only if they are the unique bulk field kinematically accesible. In the white area, DM is underabundant. One can clearly observe the resonances of the first two KK gravitons, at $m_{\rm DM}=m_1/2=210$ GeV and $m_{\rm DM}=m_2/2=384$ GeV.

\begin{figure}[htbp]
	\centering
	\includegraphics[width=0.5\textwidth]{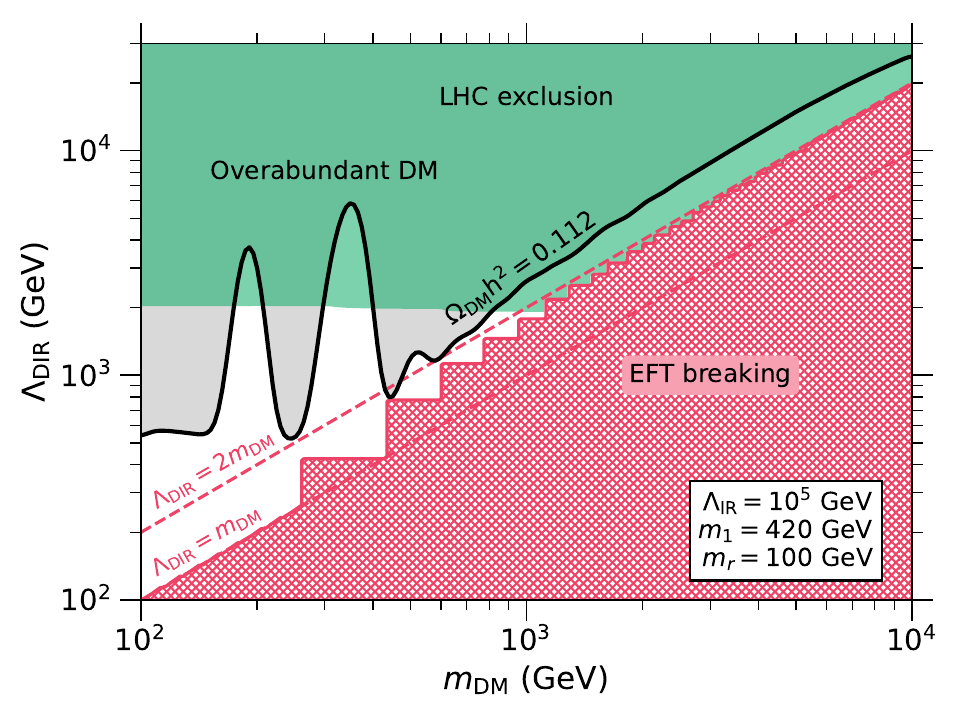}~~\includegraphics[width=0.5\textwidth]{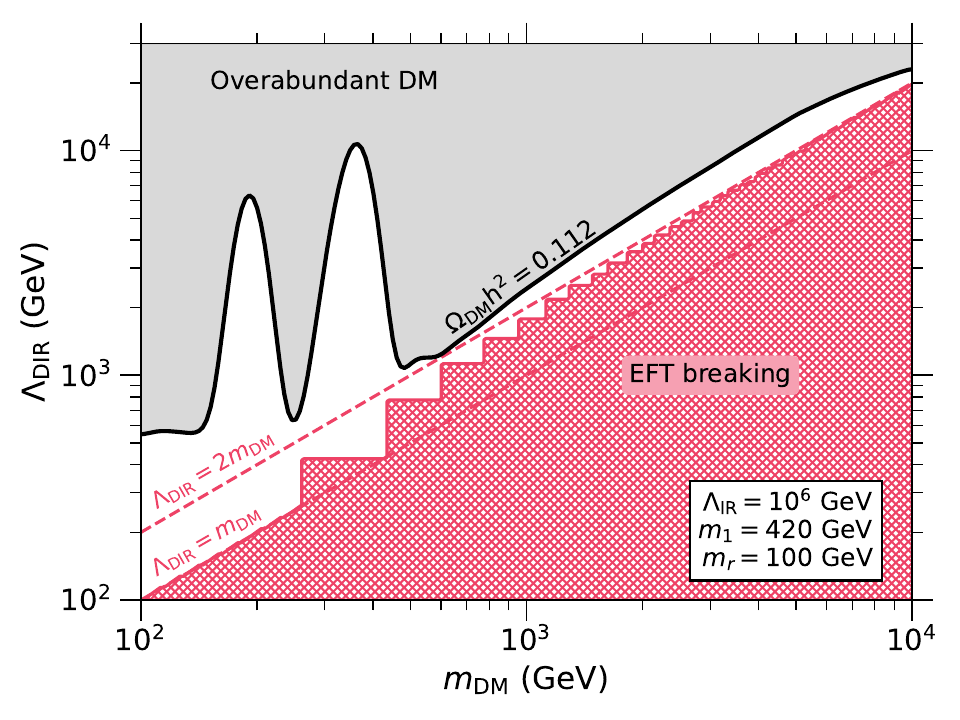}
	\caption{Results on the plane of $\Lambda_{\rm DIR}$ and the DM mass for $\Lambda_{\rm IR}=10^5$ GeV (\emph{left}) and $\Lambda_{\rm IR}=10^6$ GeV (\emph{right}). All channels (including the lightest KK gravitons) are open. The solid black line shows the relic abundance contour. } \label{fig:2Dscanall}
\end{figure}

\section{Conclusions} \label{sec:conc}

Extra-dimensional theories with enhanced gravitational interactions are a well-motivated framework to explain the DM relic abundance through thermal freeze-out. However, in the specific case of scalar DM within a two-brane RS setup, experimental constraints from LHC resonance searches rule out this possibility almost completely. In this work, we explore a scalar singlet DM embedding into a three-brane setup, where the Higgs portal is naturally forbidden. This revives the viability of the DM  production via gravitational freeze-out. Specifically, we consider a scenario where the SM resides at an IR brane, while the DM is confined to a distinct DIR brane. We consider a regime in which the values of the bulk curvatures to the left and to the right of the IR brane are similar, which makes the latter brane \emph{evanescent}. Achieving thermal equilibrium between the SM and DM sectors imposes an upper limit on the IR scale, as excessively large values suppress the interactions between the two sectors. We find that thermal freeze-out is possible for $\Lambda_{\rm IR} < 10^8$ GeV.

The interplay between the IR and DIR scales governs key phenomenological properties, including the relative rates of DM annihilation into SM particles versus radions and gravitons, as well as the constraints from DD, ID, and collider experiments. Our analysis reveals that for $\Lambda_{\rm DIR} < 10^{-2} \Lambda_{\rm IR}$, DM annihilations dominantly occur into bulk particles, thereby 
evading LHC constraints. Our numerical results indicate that for $10\,{\rm TeV} < \Lambda_{\rm IR} < 10^3$ TeV and $300\, {\rm GeV} < \Lambda_{\rm DIR} < 10$ TeV, scalar DM masses in the range $m_{\rm DM} \simeq [0.1 - 10]$ TeV are allowed, contingent on the radion and graviton mass spectrum. Some of these lower values are compatible with an RS solution to the SM electroweak hierarchy problem, see Eq.~\eqref{eq:hierarchy}. Future DD experiments may allow to probe this scenario. 

We find that our setup not only allows for a broad range of parameter choices but also that the physical implications are different. For instance, the hierarchy between $m_{\rm DM}$ and $m_1$ (with $m_{\rm DM} > m_1$) results in markedly different values of $\Lambda_{\rm DIR}$ — spanning up to two orders of magnitude — and enables the evasion of ID constraints. It is also worth emphasizing that this flexibility in parameter space has so far been explored only for scalar DM — the most constrained type in RS models. In contrast, extending the analysis to fermionic or vector DM appears to be interesting, as their properties make them inherently more viable candidates. When embedded in an \emph{evanescent} three-brane model, this framework presumably will open the door to a significantly less constrained region of parameter space, allowing different values of both $\Lambda_{\rm IR}$ and $\Lambda_{\rm DIR}$, and offering a compelling target for future DD experiments. 

Finally, let us mention that, irrespective of the DM physics, if several KK resonances are found at a collider, measurements of their mass and BRs may allow to determine the number of branes and provide a smoking gun signature of the localisation of the SM in an intermediate brane. This definitely merits further investigation.

\vspace{0.5 cm}

\section*{Acknowledgements}
We are grateful to Dipan Sengupta, Arturo de Giorgi, Roberto Ruiz de Austri and Stefan Vogl for useful discussions. This work is partially supported by the Spanish \emph{Agencia Estatal de Investigación} MICINN/AEI (10.13039/501100011033) grant PID2023-148162NB-C21, the Generalitat Valenciana through
the grant PROMETEO/2022/069, and the \emph{Severo Ochoa} project MCIU/AEI CEX2023-001292-S. JHG is supported by the \emph{Consolidación Investigadora Grant CNS2022-135592}, funded also by \emph{European Union NextGenerationEU / PRTR}, and the Generalitat Valenciana Plan GenT Excellence Program (CIESGT2024/7). GL is supported by the Generalitat Valenciana APOSTD/2023 Grant No. CIAPOS/2022/ 193. 
MGF is supported by MMT24-IFIC-01, that comes from the European Union's Recovery and Resilience Facility-Next Generation, in the framework of the General Invitation of the Spanish Government’s public business entity Red.es to participate in talent attraction and retention programmes within Investment 4 of Component 19 of the Recovery, Transformation and Resilience Plan. AMO acknowledges support from the Generalitat Valenciana programs Plan GenT Excellence Program CIDEGENT/2020/020, PROMETEO/2019/083 and CIACIF/2021/260.

\newpage

\appendix

\section{Interactions of the KK gravitons and the radion with brane-localized matter} 
\label{app:Int}

In this appendix we derive the couplings of Table~\ref{table:interactions} from the action. We define $k_1\equiv k\,, k_2=k+\delta k$. We work in the limit $\delta k/k \ll 1$, where the wave-functions of the KK gravitons and the radion formally coincide with those of the two-brane setup. In an upcoming publication~\cite{Donini:2025qrf}, we will discuss the more general case with $k_1\neq k_2$. 

Scalar and tensor perturbations over the background metric
can be added in different ways. Following Ref.~\cite{Csaki:2000zn}, we may introduce a metric
in the segment $y \in [0, L_1]$: 
\begin{equation}
\label{eq:perturbedmetric1}
ds_1^2 = e^{-2 A_1 } \left [ \left ( 1 - 2 F_1 \right )\, g_{\mu\nu} + E_{1, \mu \nu} \right ] dx^2 
- ( 1 + G_1 )^2 dy^2 \, ,
\end{equation}
and in the segment $y \in [L_1,L_2]$:
\begin{equation}
\label{eq:perturbedmetric2}
ds_2^2 = e^{-2 A_2 } \left [ \left ( 1 - 2 F_2 \right )\, g_{\mu\nu} + E_{2, \mu \nu} \right ] dx^2 
- ( 1 + G_2 )^2 dy^2 \, ,
\end{equation}
where $A_i = k_i \, y$, $g_{\mu\nu}$ is a 4D perturbed
metric (that include tensor, {\em i.e.} graviton, modes), 
and $F_i,E_i$ and $G_i$ are adimensional scalar perturbations. In the limit $\delta k/k \ll 1$, it can
be shown that the scalar perturbations $E_i$ can be gauged away. At the same time, the fields $G_i$ are not independent fluctuations, as it can be shown that
they are related to the fields $F_i$ due to the Einstein equations. Eventually, we are left with tensor perturbations of the 4D metric $g_{\mu\nu}$ and with
the scalar perturbations $F_1$ and $F_2$, each of them 
defined in one of the two segments.

The 5-dimensional graviton field can be expanded in a tower of 4-dimensional KK fields (we only focus on the spin-2 KK gravitons $h_{\mu\nu}^{(5)}(x,y)$)
\begin{equation}
    h_{\mu\nu}^{(5)}(x,y)=\sum_{n}h_{\mu\nu}^n(x)\chi^{(n)}(y)\,.
\end{equation}
The KK graviton wave-functions in the three-brane setup formally coincide with those of the two-brane setup in the limit $\delta k\to0$,
\begin{equation}
    \chi^{(0)}=\sqrt{\frac{k}{1-e^{-2kL_2}}}\simeq \sqrt{k}\,, \qquad \chi^{(n\neq0)}(y)=\frac{e^{2ky}}{N_n}J_2\left(x_n e^{k(y-L_2)}\right)\,,
\end{equation}
where $L_2\equiv \pi r_c$ and the normalization factor is given by
\begin{equation}
N_n\overset{kL_2\gg1}{\simeq}-\frac{e^{kL_2}}{\sqrt{k}}J_0(x_n)=\frac{e^{kL_2}}{\sqrt{k}}J_2(x_n)\,.
\end{equation}

As we have seen, in principle there are two (independent)
scalar fluctuations, defined to the left, $F_1(x,y)$,
and to the right, $F_2 (x,y)$, of the IR brane. In the limit $\delta k\ll k$, the scalar perturbation $F_1(x,y)$ effectively decouples from the low-energy spectrum as its mass is proportional to $1/\delta k$. On the other hand, the scalar perturbation of the metric $F_2$ in the segment $y \in [L_1,L_2]$ still 
couples with fields localized at the IR and DIR branes. It can be written in terms of the ``radion" field:
\begin{equation}
F_2(x,y) = \frac{\kappa}{2} \,  
\left (
\frac{e^{2 \left [k_2 (y - L_1) + k_1 L_1 \right ]}}{\sqrt{6}} 
\right ) 
\, {\hat r(x)} \to \frac{\kappa}{2} \,  
\left (\frac{e^{2 k y}}{\sqrt{6}} \right ) 
\, {\hat r(x)}  \, ,
\end{equation}
where $\kappa = 2/M_5^{3/2}$.
Consider now a 4D free real scalar field $\phi_0(x)$ of mass $m_0$ living in a brane localized at some generic value of the 5-dimensional coordinate $y=L$. The 5D action is given by
\begin{equation}
	    S_\phi=\frac{1}{2}\int d^5x\, \sqrt{-g^{(4)}}\,\left(\partial_\mu\phi_0\partial_\nu\phi_0 g^{\mu\nu}-m_0^2\phi_0^2\right)\delta(y-L)\,,
\end{equation}
where $g_{\mu\nu}$ is the 4D metric induced on the brane, with determinant $g^{(4)}=g^{(5)}/g^{(5)}_{55}$. In the limit $\delta k/k\ll 1$, the determinant of the induced metric and the 4D inverse metric can be expanded in powers of $\kappa$
as follows:
\begin{equation}\label{eq:Gexp1}
\sqrt{-g^{(4)}} =e^{-4 k y}
\left[1+\kappa\left(\frac{h^{(5)}}{2}-\frac{e^{2ky}}{\sqrt{6}}\hat{r}\right)\right]+\mathcal{O}(\kappa^2) \,,
\end{equation}
\begin{equation}\label{eq:Gexp2}
g^{\mu\nu}=e^{2ky}\left[\eta^{\mu\nu}+\kappa\left(-h^{(5)\,\mu\nu}+\eta^{\mu\nu}\frac{e^{2ky}}{2\sqrt{6}}\hat{r}\right)\right]+\mathcal{O}(\kappa^2)\,.
\end{equation}

We first focus on the KK gravitons couplings. 
Neglecting for the moment the radion field $\hat{r}(x)$
and expanding the metric in $\kappa$ we have:
\begin{equation}
    S_\phi \simeq \frac{1}{2}\int d^5x\, e^{-4ky}\,\left(1+\frac{\kappa}{2}h^{(5)}\right)\left[\partial_\mu\phi_0\partial_\nu\phi_0 \left(\eta^{\mu\nu}-{\kappa}h_{\mu\nu}^{(5)}\right)-m_0^2\phi_0^2\right]\delta(y-L) \, .
\end{equation}

Next, we expand the 5-dimensional graviton field $h_{\mu\nu}^{(5)}(x,y)$ in terms of its KK tower and perform the integral along the coordinate $y$. Regarding the free part of the action, we find
\begin{equation}
    S_{0,\phi}=\frac{1}{2}\int d^4x \left[(\partial_\mu\phi)^2-m^2\phi^2\right]\,,
\end{equation}
where the physical field and mass are defined as $\phi\equiv e^{-kL}\phi_0$ and $m\equiv e^{-kL}m_0$. The interaction of the $n$-th KK graviton mode with the scalar field is described by the action
\begin{equation}
	\begin{split}
   S_{\phi,h^{(n)}} &=-\frac{\kappa}{2}\int d^4x\, \chi^{(n)}(y=L)h_{\mu\nu}^n(x)\left\{\partial^\mu\phi\partial^\nu\phi-\frac{1}{2}\eta^{\mu\nu}[(\partial_\rho\phi)^2-m^2\phi^2 ]\right\}\\
   &=-\frac{\kappa}{2}\int d^4x\, \chi^n(y=L)h_{\mu\nu}^n(x)T^{\mu\nu},
    \end{split}
\end{equation}
where we recognize the expression for the energy-momentum tensor of a free real scalar field. The last equation in terms of $T_{\mu\nu}$ can be generalized to a generic matter content.

Using the explicit form of the KK gravitons wave-functions, $\kappa=2/M_5^{3/2}$ and the (approximate) relation between
the reduced Planck mass $\bar M_{\rm P}$, the fundamental scale of gravity $M_5$ and the curvature $k$, 
$M_5^3 \sim k \bar{\mpl}^2$, we get: 
\begin{equation}
   S_{\phi,h^{(n)}}= -\int d^4x \frac{h^n_{\mu\nu}T^{\mu\nu}}{\Lambda_{\rm L}^n}\,,
\end{equation}
where
\begin{equation}
\left \{
\begin{array}{l}
    \Lambda_L^0=\bar{\mpl} \, , \\
    \\
    \Lambda_L^{n\neq0}=\bar{\mpl} e^{-k(2L-L_2)}\frac{J_2(x_n)}{J_2(x_ne^{k(L-L_2)})}\overset{x_n\,e^{k(L-L_2)}\ll1}{\simeq} \bar{\mpl} e^{-k(4L-3L_2)}\frac{8J_2(x_n)}{x_n^2} \, .
    \end{array}
    \right .
\end{equation}
For $L=L_2$ we recover $\Lambda_{\rm DIR}=\bar{\mpl} e^{-kL_2}\equiv \omega\xi \bar{M_{\rm P}}$. For $L=L_1$ we get $\Lambda_{\rm IR}^n$ as in Table~\ref{table:interactions}.

The coupling of the radion can be obtained in the same way considering the 5D radion field $\hat{r}$ in the expansion of Eqs.~\eqref{eq:Gexp1} and~\eqref{eq:Gexp2}. Normalizing
properly the kinetic term for the field $\hat r$, 
in the limit $k_2 \to k_1$ we get: 
\begin{equation}
    \hat{r}(x,y)= \sqrt{k_2} \, 
    e^{- \left [ k_2 \Delta L + k_1 L_1 \right ]} \, 
    r(x) \to \sqrt{k} \, e^{- k L_2} \, r(x) \, .
\end{equation}
Eventually, we obtain the action for the interaction of the radion with matter living in a brane localized at $y=L$,
\begin{equation}
  S_{\phi,r} = \int d^4x \frac{r(x)T(x)}{\sqrt{6}\Lambda_L}\,,
\end{equation}
where $T$ is the trace of the energy-momentum tensor while
\begin{equation}
    \Lambda_L=\bar{\mpl} e^{-k(2L-L_2)}\,,
\end{equation}
which reduces to $\Lambda_{\rm DIR}$ if $L=L_2$ and to $\Lambda_{\rm IR}$ if $L=L_1$. This coupling is modified
out of the limit $\delta k/k\ll 1$ in the following way:
\begin{equation}
\Lambda_L = \bar M_{\rm P} \,  
\left ( \frac{k_2}{k_1} \right )^{1/2}
\left \{ 
\begin{array}{l}
\frac{\omega}{\xi} + {\cal O}(\omega^3) \, ,  \qquad (L = L_1) \\
\\
\omega \, \xi + {\cal O}(\omega^3) \, .  \qquad (L = L_2)
\end{array}
\right .
\end{equation}
However, out of the {\em evanescent} brane limit, it is no longer straightforward to neglect the impact of the second
radion mode, that also will couple with fields on the
IR brane. In that case, therefore, a detailed computation
that will depend on the smallness of the splitting 
$(k_2 - k_1)/k_1$ must be
carried on (see Ref.~\cite{Donini:2025qrf}).

\section{Radion-mediated dark matter direct detection}\label{app:DD}

\begin{figure}[htbp]
	\centering
	\includegraphics[width=0.5\textwidth]{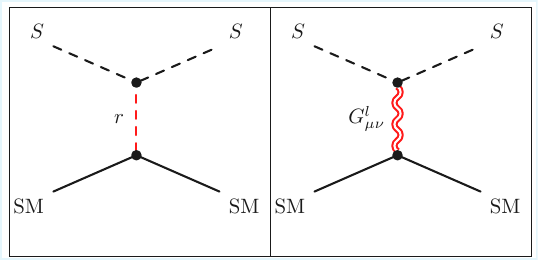}
	\caption{Direct detection diagrams for DM scattering off nucleons.} \label{fig:DDdiagrams}
\end{figure}

We consider the elastic scattering of a DM particle off a SM quark inside a nucleus mediated at tree level by a radion as shown in Fig.~\ref{fig:DDdiagrams} in the $t-$channel. The corresponding  amplitude is given by 
\begin{equation}
	\mathcal{M}_{\rm UV}=\frac{-i}{t-m_r^2}\frac{1}{6\Lambda_{\rm IR}\Lambda_{\rm DIR}}\frac{1}{4}\overline{u}_q(k_2)\left[16\,m_q-6(\slashed{k}_1+\slashed{k}_2)\right]u_q(k_1)(4\,m_{\rm DM}^2-2\,p_1\cdot p_2)\,,
\end{equation}
where $k_{1(2)}$ is the four-momentum of the quark in the initial (final) state, $p_{1(2)}$ the analogue for DM and $t=(k_1-k_2)^2$ is the usual Mandelstam variable.
In the non-relativistic limit the expression simplifies as: $t\ll m_r^2$, $\slashed{k}=k_\mu\gamma^{\mu}\simeq m_q\gamma^0$ and $p_1\cdot p_2\simeq m_{\rm DM}^2$. Furthermore, the Dirac spinor is approximately equal to $u_q\simeq \sqrt{2m_q}(\xi,0)^T$ with $\xi^{\dagger}\xi=1$ so that $\overline{u}_qu_q\simeq 2m_q$ and $\overline{u}_q\gamma^0u_q\simeq 2m_q$. Thus,
\begin{equation}
		\mathcal{M}_{\rm UV}\overset{{\rm NR}}{\simeq} \frac{4im_q^2m_{\rm DM}^2}{6m_r^2\Lambda_{\rm IR}\Lambda_{\rm DIR}}\,.
\end{equation}
In the low-energy EFT for radions and quarks, we can compute the amplitude for the elastic scattering generated by the effective scalar operator $\mathcal{O}_q=c_q^rm_q S^2\bar{q}q$, given by 
\begin{equation}
\mathcal{M}_{\mathcal{O}_q}=2ic_q^rm_q\overline{u}_q(k_2)u_q(k_1)\overset{\rm NR}{\simeq}4ic_q^rm_q^2\,.
\end{equation}
Finally, we can match the UV and the EFT results to obtain the Wilson coefficient
\begin{equation}
	c_q^r=\frac{m_{\rm DM}^2}{6m_r^2\Lambda_{\rm IR}\Lambda_{\rm DIR}}\,.
\end{equation}
Analogous computations fix the Wilson coefficients for the DM-gluon operator $\mathcal{O}_g$, as well as for the operators corresponding to the $t$-channel exchange of a KK graviton.

\begin{figure}[htbp]
	\centering	\includegraphics[width=0.7\textwidth]{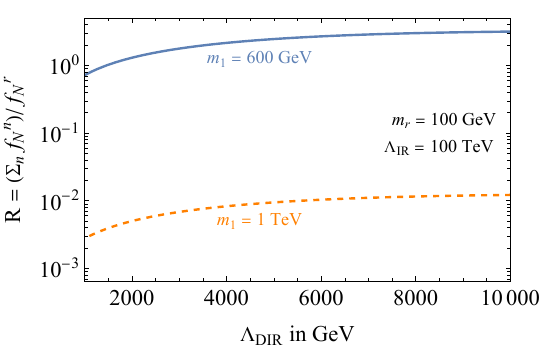}
	\caption{Ratio of the contributions of the graviton and the radion to the scattering cross-section relevant for direct detection searches. We plot the quantity $R=(\sum_{n=1}^{5}f_N^n)/f_N^r$, which quantifies the relative contribution of the first 5 KK gravitons and the radion to the cross section, versus the deep IR scale. Heavy gravitons  provide a negligible contribution while light gravitons can provide the dominant contribution.}   \label{fig:KKcouplingsb}
\end{figure}

\section{Decay widths}\label{app:decayrates}

\subsection{Radion}
The decay widths of the  lightest radion read:
\begin{equation}
\Gamma_{r \to HH} = 
\frac{
   \left(2m_H^2 + m_r^2\right)^2
}{
  192\pi \Lambda_{\mathrm{IR}}^2 m_r
} \sqrt{1 - \frac{4m_H^2}{m_r^2}}\,,
\end{equation}

\begin{equation}
\Gamma_{r \to \bar{\psi} \psi} =
\frac{N_c 
  m_r m_\psi^2 
}{
  48\pi \Lambda_{\mathrm{IR}}^2
} \left(1 - \frac{4m_\psi^2}{m_r^2}\right)^{3/2}\,,
\end{equation}

\begin{equation}
\Gamma_{r \to ZZ} = 
\frac{
  12m_Z^4 - 4m_r^2 m_Z^2 + m_r^4
}{
  192\pi \Lambda_{\mathrm{IR}}^2 m_r
} \sqrt{1 - \frac{4m_Z^2}{m_r^2}}\,,
\end{equation}

\begin{equation}
\Gamma_{r \to W^+ W^-} = 
\frac{
12m_W^4 - 4m_r^2 m_W^2 + m_r^4
}{
  96\pi \Lambda_{\mathrm{IR}}^2 m_r
} \sqrt{1 - \frac{4m_W^2}{m_r^2}}\,,
\end{equation}

\begin{equation}
\Gamma_{r \to \gamma\gamma} = 
\frac{
  C_{\mathrm{em}}^2 \alpha_{\mathrm{em}}^2 m_r^3
}{
  1536\pi^3 \Lambda_{\mathrm{IR}}^2
}\,,
\end{equation}

\begin{equation}
\Gamma_{r \to gg} = 
\frac{
  C_{\mathrm{3}}^2 \alpha_{\mathrm{s}}^2 m_r^3
}{
  192\pi^3 \Lambda_{\mathrm{IR}}^2
}\,,
\end{equation}
where $C_{\mathrm{em}}$ and  $C_{\mathrm{3}}$ are defined in Eq.~\eqref{eq:radEFT} and $\psi$ refers to SM fermions.

\subsection{KK graviton}

The decay widths of the  KK gravitons read:
\begin{equation}
\Gamma_{G_n \to HH} = 
\frac{
  \left(m_n^2 - 4m_H^2\right)^2
}{
  960\pi \left(\Lambda_\mathrm{IR}^n\right)^2 m_n
}  \sqrt{1 - \frac{4m_H^2}{m_n^2}}\,,
\end{equation}

\begin{equation}
\Gamma_{G_n \to \bar{\psi} \psi} = 
\frac{ N_c
  m_n^3 
}{
  160\pi \left(\Lambda_\mathrm{IR}^n\right)^2
} \left(1 - \frac{4m_\psi^2}{m_n^2}\right)^{3/2} \left(1 + \frac{8m_\psi^2}{3m_n^2}\right)\,,
\end{equation}

\begin{equation}
\Gamma_{G_n \to ZZ} = 
\frac{
   \left(56m_n^2 m_W^2 + 13m_n^4 + 48m_Z^4\right)
}{
  960\pi \left(\Lambda_\mathrm{IR}^n\right)^2 m_n
} \sqrt{1 - \frac{4m_Z^2}{m_n^2}}\,,
\end{equation}

\begin{equation}
\Gamma_{G_n \to W^+ W^-} = 
\frac{
   \left(56m_n^2 m_W^2 + 13m_n^4 + 48m_Z^4\right)
}{
  480\pi \left(\Lambda_\mathrm{IR}^n\right)^2 m_n
} \sqrt{1 - \frac{4m_Z^2}{m_n^2}}\,,
\end{equation}

\begin{equation}
\Gamma_{G_n \to \gamma\gamma} = 
\frac{
  m_n^3
}{
  80\pi \left(\Lambda_\mathrm{IR}^n\right)^2
}\,,
\end{equation}

\begin{equation}
\Gamma_{G_n \to gg} = 
\frac{
  m_n^3
}{
  10\pi \left(\Lambda_\mathrm{IR}^n\right)^2
}\,,
\end{equation}

\begin{equation}
\Gamma_{G_n \to \mathrm{SM}} \approx 
\frac{73 m_n^3}{240\pi \left(\Lambda_\mathrm{IR}^n\right)^2} \,,
\end{equation}

\begin{equation}
\Gamma_{G_n \to \mathrm{rr}} = 
\frac{\chi_{nrr}^2\left(m_n^2-4m_r^2\right)^{5/2}}{960\pi\Lambda_\mathrm{DIR}^2m_n^2}\,,
\end{equation}

\begin{align}
\Gamma_{G_n \to rG_m} &= 
\frac{
    \left(k^4 e^{-4kL_2\pi}\right) \tilde{\chi}_{rnm}^2\sqrt{Q\left(m_m,m_n,m_r\right)}
}{
    480\pi \Lambda_\mathrm{DIR}^2m_n^7 m_m^4
}
    \notag \\
    &\quad \times\Bigg(
        2m_m^6(13m_n^2 - 2m_r^2)
        + 2m_m^4(-28m_n^2 m_r^2 + 63m_n^4 + 3m_r^4)  \notag \\
    &\quad 
        +\, 2m_m^2(13m_n^2 - 2m_r^2)(m_n^2 - m_r^2)^2
        + m_m^8 + (m_n^2 - m_r^2)^4
\Bigg)\,,
\end{align}

\begin{equation}
\Gamma_{G_n \to \mathrm{G_m G_m}} = 
\frac{\chi_{nmm}^2\left(m_n^2-4m_m^2\right)^{5/2}\left(780m_m^6m_n^2+616m_m^4m_n^4+52m_m^2m_n^6+180m_m^8+m_n^8\right)}{34560\pi\Lambda_\mathrm{DIR}^2m_m^8m_n^2}\,,
\end{equation}

\begin{align}
\Gamma_{G_n \to G_mG_k} &= 
\frac{
    \chi_{nmk}^2Q^{5/2}\left(m_k,m_m,m_n\right)
}{
    17280\pi \Lambda_\mathrm{DIR}^2m_k^4 m_m^4 m_n^7 
}
\Bigg(26m_k^2(14m_m^4 m_n^2 + 14m_m^2 m_n^4+ m_m^6 + m_n^6)
    \notag \\
    &\quad
        26m_k^6(m_m^2 + m_n^2)
        + 14m_k^4(26m_m^2 m_n^2 + 9m_m^4 + 9m_n^4)
         \notag \\      
    &\quad
        +\, m_k^8 + 26m_m^2 m_n^2 + 126m_m^4 m_n^4 + 26m_m^6 m_n^2
        + m_m^8 + m_n^8
\Bigg)\,,
\end{align}
where $\psi$ refers to SM fermions. Here $\Lambda^n_{\rm IR}$ is the scale associated to the $n$-KK graviton, defined in Eq.~\eqref{eq:LambdaIR}. We also make use of the factors:
\begin{equation}
\chi_{nrr} \equiv 
\frac{-2}{J_0\left(x_n\right)}\int_{0}^{1}du\,\,  u^3J_2\left(x_n u\right)\,,
\end{equation}
\begin{equation}
\tilde{\chi}_{rnm} \equiv 
2\frac{x_nx_m}{J_0\left(x_n\right)J_0\left(x_m\right)}\int_{0}^{1}du\,\, u^3J_1\left(x_n u\right)J_1\left(x_m u\right)\,,
\end{equation}
\begin{equation}
\chi_{nmk} \equiv 
\frac{-2}{J_0\left(x_n\right)J_0\left(x_m\right)J_0\left(x_k\right)}\int_{0}^{1}du\,\, u^3J_2\left(x_n u\right)J_2\left(x_m u\right)J_2\left(x_k u\right)\,,
\end{equation}
and the function:
\begin{equation}
Q\left(x,y,z\right)  =
\left(x-y-z\right)\left(x+y-z\right)\left(x-y+z\right)\left(x+y+z\right)\,.
\end{equation}

\bibliographystyle{JHEP}
\bibliography{paper}

\end{document}